\documentclass[a4paper,onecolumn,11pt,accepted=2026-09-09]{quantumarticle}
\pdfoutput=1
\usepackage[utf8]{inputenc}
\usepackage[english]{babel}
\usepackage[T1]{fontenc}

\usepackage{graphicx}% Include figure files
\usepackage{dcolumn}% Align table columns on decimal point
\usepackage{bm}% bold math
\usepackage{siunitx}
\usepackage{physics}
\usepackage{xcolor}
\usepackage{booktabs}
\usepackage{colortbl}
\usepackage[caption=false]{subfig}
\usepackage[export]{adjustbox}
\usepackage{comment}
\usepackage{amssymb}

\usepackage{hyperref}
\hypersetup{
        colorlinks=true
}

\usepackage{nicematrix}
\NiceMatrixOptions{
code-for-first-row = \color{black} ,
code-for-last-row = \color{black} ,
code-for-first-col = \color{black} ,
code-for-last-col = \color{black}
}

\newcommand{\aop}{\hat{a}}
\newcommand{\adag}{\hat{a}^\dagger}
\newcommand{\adagT}{\hat{a}^{\dagger 2}}

\newcommand{\adagn}{\hat{a}^{\dagger n}}
\newcommand{\sigP}{\hat{\sigma}_+}
\newcommand{\sigM}{\hat{\sigma}_-}
\newcommand{\sigZ}{\hat{\sigma}_z}
\newcommand{\sigX}{\hat{\sigma}_x}
\newcommand{\sigY}{\hat{\sigma}_y}

\begin{document}

\title{Fast Bosonic Control via Multiphoton Qubit-Oscillator Interactions}

\author{Noah Gorgichuk}

\affiliation{%
 Institute for Quantum Computing, University of Waterloo, 200 University Avenue West, Waterloo,
Ontario N2L 3G1, Canada}
\affiliation{Department of Physics and Astronomy, University of Waterloo,
200 University Avenue West, Waterloo, Ontario N2L 3G1, Canada}
\affiliation{Red Blue Quantum Inc., 72 Ellis Crescent North, Waterloo, Ontario N2J 3N8, Canada}

\author{Mohammad Ayyash}
\thanks{Corresponding author: mmayyash@protonmail.com}

\affiliation{Red Blue Quantum Inc., 72 Ellis Crescent North, Waterloo, Ontario N2J 3N8, Canada}

\author{Matteo Mariantoni}

\affiliation{%
 Institute for Quantum Computing, University of Waterloo, 200 University Avenue West, Waterloo,
Ontario N2L 3G1, Canada}
\affiliation{Department of Physics and Astronomy, University of Waterloo,
200 University Avenue West, Waterloo, Ontario N2L 3G1, Canada}
\affiliation{Red Blue Quantum Inc., 72 Ellis Crescent North, Waterloo, Ontario N2J 3N8, Canada}
\author{Sahel Ashhab}

\affiliation{Advanced ICT Research Institute, National Institute of Information and
Communications Technology, 4-2-1, Nukui-Kitamachi, Koganei, Tokyo 184-8795, Japan}
\affiliation{Research Institute for Science and Technology, Tokyo University of Science, 1-3 Kagurazaka, Shinjuku-ku, Tokyo 162-8601, Japan}
\affiliation{Department of Physics, The University of Tokyo, Tokyo 113-0033, Japan}
    
\maketitle
   
\begin{abstract}
    We study the problem of $n$-fold rotationally symmetric bosonic state preparation, which is of great importance to bosonic quantum error correction, using a multiphoton interaction between an oscillator and an auxiliary qubit. We present an $n$-photon Law-Eberly ($n$LE) protocol that serves as an analytic baseline, alongside numerical optimal control calculations that further reduce state preparation time. We find that multiphoton control protocols substantially reduce preparation times for binomial, cat, and Gottesman-Kitaev-Preskill codewords compared to schemes relying on standard linear interactions. Further, we achieve arbitrary control over the oscillator's Hilbert space by combining different multiphoton interaction orders. We also extend these control improvements to the preparation of rotationally symmetric multi-oscillator states. Lastly, numerical simulations using realistic planar superconducting circuit parameters validate the robustness of our scheme against qubit and oscillator decoherence. Our findings can significantly enhance the performance of bosonic codes on planar superconducting hardware, an important ingredient for scalable fault-tolerant quantum computers.
\end{abstract}

\section{Introduction}

Bosonic quantum computation leverages the infinite-dimensional Hilbert space of a harmonic oscillator to protect quantum information. The protection in typical codes is associated with symmetries of the codespace such as invariance under certain rotations \cite{Grimsmo_RotSymm_BosonicQEC_2020} or displacements \cite{Gottesman_EncQubOsc_2001}. Bosonic codes can be implemented by synthesizing a codeword state in an oscillator (e.g., direct generation of Gottesman–Kitaev–Preskill (GKP) states \cite{Campagne-Ibarcq_QECGrid_CondDisp_2020,Eickbusch_FastUnivCont_Disp2022}). It is also possible to engineer a dynamical system such that the codewords arise as eigenstates of its Hamiltonian (e.g., two-photon driven Kerr parametric oscillators \cite{Puri_StabCat_2019,Grimm2020stabilization,Iyama_ObservationKPO_2024}) or as steady states of its Lindbladian dissipator (e.g., dissipatively stabilized cat states \cite{Mirrahimi_DynProt_CatQubits_2014,Leghtas_TwoPhotonDissip_CatExp_2015,Vaneslow_FourPhotonDissip_2025}). Arbitrary control over the oscillator's Hilbert space is necessary for manual synthesis of logical codewords. Such control can be facilitated by cubic (or higher-order) oscillator Hamiltonians \cite{Lloyd_QCCV_1999,Sefi_DecompCVQO_2011} or by interactions with an ancillary qubit \cite{LawEberely_ArbControl_1996,Krastanov_SNAP_2015,Strauch_AllResControl_2012,Campagne-Ibarcq_QECGrid_CondDisp_2020}.

Previous demonstrations of bosonic codeword synthesis on state-of-the-art superconducting hardware have been restricted to oscillators encoded in electromagnetic field modes of three-dimensional cavities \cite{Eickbusch_FastUnivCont_Disp2022,Sivak_RealTimeQEC_2023}. This is because of the long state preparation times needed to populate the oscillator with many photons as prescribed by the particular control scheme \cite{Eickbusch_FastUnivCont_Disp2022}. The coherence times of planar superconducting resonators coupled to nonlinear elements are still much lower than those of three-dimensional cavities \cite{Blais_cQEDReview_2021}. This currently stands as a major bottleneck in the scalability of bosonic codes in superconducting circuits, as the transition from three-dimensional to planar resonators is an important and almost inevitable step towards scalable architectures. Thus, to improve the performance of bosonic codes on planar circuits, it is necessary to improve the device coherence times and for new control methods to shorten the state preparation and gate operation times.

Most state-of-the-art demonstrations of bosonic codeword preparation have relied on linear interactions between a qubit and an oscillator in superconducting circuit \cite{Campagne-Ibarcq_QECGrid_CondDisp_2020,Eickbusch_FastUnivCont_Disp2022,Sivak_RealTimeQEC_2023} and ion trap \cite{Fluhmann_QECGrid_TrappedIons_2019,Matsos_GKPTrappedIonsExp_2025} platforms. Although linear qubit-oscillator interactions still enable nonlinear phenomena such as two-photon transitions \cite{Deppe_TwoPhotonProbJC_2008,Ourjoumtsev_AtomicTwoPhotonTrans_2011}, more recently, however, nonlinear interactions creating multiple simultaneous excitations in an oscillator have been demonstrated experimentally in superconducting circuits \cite{Sandbo_ThreePhotonSPDC_2020,Eriksson_UnivControl_Cubic_2024}. As a result, a number of studies exploring the use of nonlinear qubit-oscillator interactions in control \cite{Ayyash_DrivenTwoPhoton_2024,Hope_ConditionalSqueezing_2025}, readout \cite{Chapple_BalancedCrossK_Readout_2025} and other quantum information processing applications have emerged \cite{Anai_CtrlSqzProjection_2024,Ayyash_DrivenTwoPhoton_2024,Tang_TwoPh_CZ_2025}. 

In this paper, we propose the use of nonlinear qubit-oscillator interactions combined with a qubit drive for preparing multiphoton states in the oscillator, specifically bosonic codewords. The state preparation relies on populating the oscillator with bundles of photons rather than one photon at a time, reducing the number of operations needed. This serves as a multiphoton extension of the Law and Eberly protocol \cite{LawEberely_ArbControl_1996}, which was the first arbitrary oscillator control protocol relying on linear qubit-oscillator interactions. More importantly, we show that despite the weaker coupling strengths of higher-order interactions, the combination of fewer required operations and the faster scaling of excitation swap frequencies in the multiphoton regime can result in faster state preparation. This new protocol can improve the performance of bosonic codes on planar superconducting hardware thanks to the reduced state preparation time. The multiphoton interactions required are difficult to implement and have limitations, e.g. current technology imposes a maximum Fock cutoff on which these interactions can achieve high-fidelity operations. We discuss the limitations in detail and explain in which cases the substantial improvements are robust.

The paper is structured as follows. Section~\ref{sec:Prelims} briefly reviews control methods based on linear qubit-oscillator interactions and discusses the symmetries of some bosonic codewords. In Sec.~\ref{sec:Multi_control}, we lay out the multiphoton control scheme and present examples of state preparation including multi-component cat and GKP states. Additionally, we compare the linear and $n$-photon LE protocols to conditional displacement control schemes. Section~\ref{sec:ArbControl} outlines a protocol for arbitrary control involving different orders of multiphoton interactions. We generalize the multiphoton control to the case of multiple oscillators in Sec.~\ref{sec:Multimode}. Section~\ref{sec:cQED} details a superconducting circuit quantum electrodynamics (QED) implementation with realistic parameter estimates including decoherence as well as practical limitations of the introduced protocols. Lastly, we provide our conclusions and an outlook in Sec.~\ref{sec:Conc}.

\section{Preliminaries}\label{sec:Prelims}

In this section, we briefly review some common control schemes that rely on an auxiliary qubit coupled to an oscillator. We also highlight some symmetries of bosonic codes that we later leverage in state preparation using our protocol.

\subsection{Control via linear qubit-oscillator interactions}

\subsubsection{Law-Eberly scheme}
The Law-Eberly (LE) scheme \cite{LawEberely_ArbControl_1996} makes use of a qubit drive and a single-photon Jaynes-Cummings (JC) interaction between the qubit and the oscillator. This protocol is a special case of our $n$-photon LE protocol, introduced later in this paper, for $n=1.$ The drive and JC interaction are never ``switched on" at the same time. In time steps where the resonant qubit drive is activated (the drive frequency is set to $\omega_d=\omega_q$), the system Hamiltonian in the interaction picture reads
\begin{align}\label{eq:QubitDrive}
    \hat{H}=\Omega(t)\sigP + \Omega^*(t)\sigM,
\end{align}
where $\Omega(t)$ is the qubit drive strength. Conversely, when the resonant JC interaction is activated, the interaction picture Hamiltonian is
\begin{align}
    \hat{H}=g(t)\sigP\aop + g^*(t)\sigM\adag,
\end{align}
with $g(t)$ being the linear qubit-oscillator interaction strength. The qubit drive allows for arbitrary qubit rotations, while the JC interaction allows for qubit-oscillator excitation swaps, $\ket{e}\ket{l}\leftrightarrow\ket{g}\ket{l+1}$. The LE scheme is a sequence of $2M$ operations that takes a system initialized in $\ket{g}\ket{0}$ to 
\begin{align}
    \ket{g}\ket{\Psi_M}=\ket{g}\left(\sum_{l=0}^Mc_l\ket{l}\right),
\end{align}
where $\ket{\Psi_M}$ is a target state with maximum Fock support ending at $\ket{M}$ and $\{c_l\}$ can be any set of complex numbers that obey $\sum_{l=0}^M|c_l|^2=1.$ The qubit drive is applied followed by the JC interaction, and these pairs of operations are repeated $M$ times (yielding $2M$ operations). The target state imposes constraints on the coupling and drive strengths ($g(t)$ and $\Omega(t)$), as well as the length of each time step needed to synthesize the coefficients $\{c_l\}.$ Notably, the LE scheme imposes the condition that at the end of each pair of operations (qubit drive followed by a JC interaction) the largest Fock state is only linked to the qubit ground state. This scheme has been experimentally implemented in superconducting circuits \cite{Hofheinz_SynthesisExp_2009}. In this paper, we generalize the LE scheme to use a multiphoton JC interaction that can perform qubit-oscillator swaps $\ket{e}\ket{l}\leftrightarrow\ket{g}\ket{l+n}$ with $n$ being the interaction order. This can reduce the state preparation times, as performing such swaps with a linear interaction requires more steps. We can recover the constraint equations required for the LE scheme from our more general protocol presented in Sec.~\ref{sec:Multi_control} for the case of $n=1$.

\subsubsection{SNAP scheme}
The selective number-dependent arbitrary phase (SNAP) scheme \cite{Krastanov_SNAP_2015, Heeres_SNAP_Exp_2015} relies on the JC interaction in the dispersive regime, where the effective interaction picture Hamiltonian is
\begin{align}\label{eq:DispersiveHam}
    \hat{H}_{\text{disp}}=\frac{\chi}{2}\sigZ\adag\aop.
\end{align}
Here, $\chi=g^2/\Delta$ with $\Delta=\omega_q-\omega_o$ being the detuning between the qubit and oscillator. A sequence of qubit drive pulses in the presence of the dispersive interaction realizes the SNAP operation
\begin{align}
    \hat{S}_n(\Theta)=e^{i\Theta\dyad{n}};
\end{align}
this operation applies a phase shift of angle $\Theta$ to a specific Fock state. The SNAP operation, together with displacement, $\hat{D}(\alpha)=\exp(\alpha \adag - \alpha^*\aop)$ with $\alpha\in\mathbb{C}$, allows the synthesis of arbitrary states in the oscillator \cite{Krastanov_SNAP_2015}. Here, the displacement is generated by an oscillator driving Hamiltonian,
\begin{align}
    \hat{H}=\widetilde{\Omega}(t)\adag+\widetilde{\Omega}^*(t)\aop,
\end{align}
where $\widetilde{\Omega}(t)$ is the oscillator driving strength.

\subsubsection{Conditional displacements scheme}
The third scheme relies on Eq.~\eqref{eq:DispersiveHam} in the presence of an oscillator drive to generate qubit-conditional displacements \cite{Campagne-Ibarcq_QECGrid_CondDisp_2020,Eickbusch_FastUnivCont_Disp2022},
\begin{align}\label{eq:ConditionalDisplacement}
    \hat{U}_{\text{CD}}(\alpha)=\dyad{g}\hat{D}(\alpha)+\dyad{e}\hat{D}(-\alpha).
\end{align}
It is worth noting that conditional displacements can also be generated without resorting to the dispersive regime by driving the qubit cross-resonantly at the oscillator frequency \cite{Touzard_GateConditionalDisp_2019,Ayyash_ResSchCats_2024}. Conditional displacements combined with qubit rotations allow for arbitrary state synthesis \cite{Campagne-Ibarcq_QECGrid_CondDisp_2020}.

It is worth noting that using qubit rotations with either a JC interaction or conditional displacements allows for arbitrary state preparation in the joint qubit-oscillator Hilbert space, whereas using the SNAP scheme can only prepare arbitrary states in the oscillator \cite{Liu_HybridQOProc_2024}. The JC interaction along with qubit rotations can be viewed as a \textit{sideband instruction set} by rotating the qubit and populating the oscillator Fock-state-by-Fock-state. This is due to the JC interaction being viewed as `red sideband' interaction and the anti-JC interaction, $\sigP\adag+\sigM\aop$, being viewed as a `blue sideband' interaction.  Conditional displacements combined with qubit rotations can be viewed as a \textit{phase-space instruction set} by rotating the qubit and populating the oscillator pixel-by-pixel in phase space\footnote{Pictorially, the Wigner function of a coherent (displaced vacuum) state is a blob in phase space, which can be seen as a pixel. Conditional displacements superpose many displacements together, thus populating many pixels together.}.

\subsection{Bosonic code symmetries}

Bosonic codewords are typically constructed from a subspace of the infinite-dimensional oscillator Hilbert space. In the case of rotationally symmetric codes, i.e.,~ones where the code possesses $n$-fold rotational symmetry or, equivalently, invariance under $\hat{R}(2\pi/n)=\exp(i(2\pi/n)\adag\aop)$, the codewords have exclusive Fock support on states $\{\ket{ln+k}\}_{l=0}^\infty$ for some $k\in\{0,1,...,n-1\}$~\cite{Grimsmo_RotSymm_BosonicQEC_2020}. Here, we define the $n$-photon multiples subspace shifted by $k$ photons as
\begin{align}
    \mathcal{H}_{k,n}=\text{span}\{\ket{k},\ket{n+k},\ket{2n+k},\ket{3n+k},...\}.
\end{align}
In this sense, the total oscillator Hilbert space is segmented into a direct sum of these subspaces,
\begin{align}
\mathcal{H}_{\text{total}}=\bigoplus_{k=0}^{n-1}\mathcal{H}_{k,n}.
\end{align}
Note that two oscillator states, $\ket{\psi_{k_1}}$ and $\ket{\psi_{k_2}}$, possessing exclusive Fock support on $\mathcal{H}_{k_1,n}$ and $\mathcal{H}_{k_2,n}$ ($k_1\neq k_2$), respectively, are orthogonal. This orthogonality enables the detection and correction of photon losses using rotationally symmetric bosonic codes \cite{Grimsmo_RotSymm_BosonicQEC_2020}. Although GKP codes are by definition translationally-invariant, GKP qubit states also posses a two-fold rotational symmetry and both codewords belong to $\mathcal{H}_{0,2}$ \cite{Gottesman_EncQubOsc_2001,Royer_EncodMultimodeGrid_2022}. This is because every state possessing $n$-fold rotational symmetry must belong to $\mathcal{H}_{k,n}$ for some $k$. Other codes that possess $n$-fold rotational symmetry include cat codes \cite{Cochrane_CatCode_1999,Bergmann_QECMultiCompCat_2016} and binomial codes \cite{Michael_BinomialCode_2016}. For example, the two-component and four-component cat codes are invariant under $\pi$ (two-fold)  and $\pi/2$ (four-fold) rotations, respectively.

This segmentation of the Hilbert space based on rotational symmetry (equivalently, exclusive support on $n$-photon multiples) is straightforwardly extendable to multiple oscillators.  In the case of a simple concatenation of two oscillators, the two-oscillator Hilbert space can be expressed as 
\begin{align}
\mathcal{H}_{\text{total}}=\mathcal{H}_{1}\otimes\mathcal{H}_{2}=\bigoplus_{k_1=0}^{n_1-1}\bigoplus_{k_2=0}^{n_2-1}\mathcal{H}_{k_1,n_1}\otimes \mathcal{H}_{k_2,n_2}.
\end{align}
This can be part of a concatenation with some additional layer of quantum error correction using, for example, surface codes \cite{Grimsmo_RotSymm_BosonicQEC_2020,Noh_Surface-GKP_2020}, or part of an intrinsically multimode rotationally symmetric code \cite{Ahmed_MultimodeRotSymm_2025}.

In what follows, our protocol will take advantage of the segmentation of the Hilbert space and enable fast state preparation in the subspaces $\mathcal{H}_{k,n}.$

\section{Rotationally Symmetric State Preparation via Multiphoton Control}\label{sec:Multi_control}

In this section, we outline the $n$LE protocol for generating arbitrary states that have exclusive Fock support on $\mathcal{H}_{0,n}$. This is a form of nonlinear sideband control. We then showcase several examples how the protocol works for preparing oscillator states that serve as logical states in specific bosonic encodings. In particular, we compare the preparation of these states using $n$LE with synthesis using LE employing a linear interaction as well as conditional displacement control. Additionally, we perform optimal control calculations that provide speedups to the $n$LE protocol.

\subsection{$n$-photon Law-Eberly protocol}\label{sec:Multi_control_basic_prot}
We begin by describing the generalized $n$-photon Law-Eberly ($n$LE) protocol. Then, we examine the time scaling for different interaction orders due to the multiphoton Rabi swaps and its interplay with the strength of the qubit drive.  We then consider the analytic solutions for a few example states.

\subsubsection{Protocol description}
Consider a qubit-oscillator system in the presence of a qubit drive and an $n$-photon JC ($n$JC) interaction between the qubit and oscillator. When the qubit and oscillator are near $n$-photon resonance ($\omega_q=n\omega_o$), the system Hamiltonian in the interaction picture reads
\begin{align}
    \hat{H}^{I,n}=g_n(t)\sigP\aop^n+g_n^*(t)\sigM\adagn,
\end{align}
where $g_n(t)$ is the $n$-photon interaction coupling strength. The qubit drive is modeled by the Hamiltonian of Eq.~\eqref{eq:QubitDrive}. In a single time step $\tau_j$, only one of the two interactions occurs, that is, if $g_n(\tau_j)\neq0$ then $\Omega(\tau_j)=0$ and vice versa. Without loss of generality, we simply assume the $n$JC coupling and qubit drive strength to be fixed and the same for all time steps they are switched on, and we  simply denote them by $g_n$ and $\Omega$, respectively. Here, we denote interaction time steps by $\tau_j$ and driving time steps by $\tau_j'.$ We denote the time-evolution operator for a single time step where there is only a qubit drive as
\begin{equation}
    \hat{C}_{j} = 
    \begin{pmatrix}
        \cos(|\Omega|\tau_j' ) & -ie^{i\theta }\sin(|\Omega|\tau _j')  \\
        \\
        -ie^{-i\theta} \sin(|\Omega|\tau_j' ) & \cos(|\Omega|\tau _j')
    \end{pmatrix},
\end{equation}
where $\theta=\arg(\Omega).$ Notably, we do not require qubit rotations about the qubit $z$-axis for this protocol. The time-evolution operator for a time step where only the $n$JC interaction is activated is denoted as a $2\times2$ matrix in the qubit basis $\{\ket{e},\ket{g}\},$  \cite{Phoenix_Knight_JCTimeEvOp}
\begin{align}
    \hat{Q}_{j}^{(n)} = 
    \begin{pmatrix}
        \cos(|g_{n}|\tau_j \sqrt{\hat{a}^{n} \hat{a}^{\dagger n}}) & -ie^{i\phi} \aop^n \hat{S}_\text{N}(n,j) \\
        -ie^{-i\phi} \aop^{\dagger n}\hat{S}_\text{A}(n,j)& \cos(|g_{n}|\tau_j \sqrt{\hat{a}^{\dagger n} \hat{a}^{n}})
    \end{pmatrix}, \label{eq:Q_matrix}
\end{align}
where
\begin{subequations}
    \begin{align}
        \hat{S}_\text{N}(n,j)=\frac{\sin(|g_{n}|\tau_j \sqrt{\hat{a}^{\dagger n} \hat{a}^{ n}}) }{\sqrt{\hat{a}^{\dagger n} \hat{a}^{ n}}},
    \end{align}
    \begin{align}
        \hat{S}_\text{A}(n,j)=\frac{\sin(|g_{n}|\tau_j \sqrt{\hat{a}^{n} \hat{a}^{\dagger n}})}{\sqrt{\hat{a}^{n} \hat{a}^{\dagger n}}}, 
    \end{align}
\end{subequations}
and $\phi_j=\arg(g_{n}).$ The form of Eq.~\eqref{eq:Q_matrix} is a straightforward generalization of the matrix form of the linear JC time-evolution operator \cite{Phoenix_Knight_JCTimeEvOp}. When $\hat{Q}_j^{(n)}$ is applied to a given state $\ket{g/e}\ket{l}$, the diagonal elements correspond to the operator contributions that preserve the state, up to a phase. The off-diagonal elements, in their normal and anti-normal order forms,  $\hat{S}_{\text{N}}$ and $\hat{S}_\text{A}$, respectively, are responsible for multiphoton swaps between the qubit and oscillator.

We are interested in using the $n$JC interaction along with the qubit drive to generate states that possess support exclusively on $\mathcal{H}_{0,n}.$ In other words, we can synthesize a target state of the form $\ket{\Psi_t}=\sum_{l=0}^M c_l\ket{ln}$. This target state can be truncated version of a state with infinite support over $\mathcal{H}_{0,n},$ i.e. $\ket{\Psi_t}=P_{0,n}^{(M)}\ket{\psi_t}/||P_{0,n}^{(M)}\ket{\psi_t}||,$ where $\ket{\psi_t}=\sum_{l=0}^\infty \widetilde{c}_l\ket{ln}$ and $P_{k,n}^{(M)}=\sum_{l=0}^M\dyad{ln+k}$. By selecting a sufficiently large $M$, high-fidelity approximations of a target state can be obtained, as the coefficients with higher indices contribute less. Following LE \cite{LawEberely_ArbControl_1996}, to synthesize the target state starting from an initial state $\ket{g,0}$, we consider a sequence of $2M$ operations. In this sequence, we apply a qubit drive for a single time step followed by an $n$JC interaction in the following time step, where we repeat these pairs of operations. The total unitary sequence reads
\begin{align}
    \hat{U}_t=\prod_{j=1}^{M}\hat{Q}^{(n)}_{j}\hat{C}_j,
\end{align}
where the target state is obtained as $\ket{\Psi_{t}}=\hat{U}_t\ket {g,0}$. Conceptually, the role of the qubit drive in the $k$th step is to excite the qubit state producing $\ket{e, (k-1)n}$, while the $n$JC interaction transfers part of this population to $\ket{g,kn},$ as needed for the target state. Hence, we can build a state by populating higher excitations through pairs of qubit drives and $n$JC interactions. The coupling coefficients necessary to synthesize the target state can be easily found by inverting the target state generation procedure, i.e., $\ket{g,0}=\hat{U}^\dagger_t\ket{\Psi_{t}}$. This inversion allows one to solve for $g_{n}\tau_k$ and $\Omega\tau_k'$ at each step by sequentially depopulating the highest occupied state without concern for the effect of the operation on lower states, since they will be systematically removed in subsequent steps. The sequence of coupling parameters needed for the state synthesis is obtained by the iterative removal of excitations until the initial state $\ket{g,0}$ is recovered. This is the same procedure as in the LE protocol but with $n$-photon qubit-oscillator interactions. 

To extract the exact parameters at each step, we consider subsequences of $\hat{U}_t$, and we use the intermediate state resulting from these subsequences. We define the $k$th intermediate state as
\begin{align}
    \ket{\Psi_{k}}=\prod_{j=1}^k\hat{Q}_j^{(n)}\hat{C}_j\ket{g,0},
\end{align}
where the largest Fock state occupation in the $k$th intermediate state is $\ket{kn}.$ We impose that at the end of each intermediate step the highest Fock state, $\ket{kn}$, is only linked to the ground state of the qubit, i.e., the amplitude of $\ket{e,kn}$ is zero. We now describe how the inversion algorithm works using this intermediate state. It can be assumed that the $k$th intermediate state has the explicit form
\begin{align}\label{eq:IntermediateState}
    \ket{\Psi_{k}}=c_{g,k}\ket{g,kn}+\sum_{l=0}^{k-1}(c_{g,l}\ket{g,ln}+c_{e,l}\ket{e,ln}).
\end{align}
Note that the largest Fock state is linked only to the ground state of the qubit. First, we invert the $k$th $n$JC interaction by applying $\hat{Q}_{k}^{(n)\dagger}$,
\begin{align}\label{eq:InvertedEq}
    \hat{Q}_{k}^{(n)\dagger}\ket{\Psi_{k}}=& (c_{g,k}\alpha_{k,k} + ic_{e,k-1}\beta_{k-1,k})\ket{g,kn}\nonumber\\ &+\sum_{l=0}^{k-1}(\widetilde{c}_{g,l}\ket{g,ln}+\widetilde{c}_{e,l}\ket{e,ln}) 
\end{align}
where $\alpha_{k,k}=\cos(|g_{n}|\tau_k \xi(kn,n)),$ $\beta_{k-1,k}=e^{-i\phi} \sin(|g_{n}|\tau_k \xi(kn,n))$, and 
\begin{align}
    \xi(a,b)
    = \begin{cases} 
            \sqrt{\frac{a!}{(a-b)!}} & \text{if  } a \geq b \\[0.3cm]
            0 & \text{if  } a < b\nonumber
        \end{cases}.\\ 
\end{align}
Moreover, we have
$$\widetilde{c}_{g/e,l} = \langle g/e,l | \hat{Q}_{k}^{(n)\dagger} | \Psi_{k} \rangle.
$$
Given the form of Eq.~\eqref{eq:InvertedEq}, we seek to eliminate the amplitude of $\ket{g,kn}$ such that the largest Fock state contribution is removed. We are essentially depopulating the highest Fock levels and inverting for the $n$JC swap time needed to perfectly depopulate this Fock state. This entails solving the constraint equation
\begin{align}\label{eq:JCConst}
    c_{g,k}\cos(|g_{n}|\tau_k \xi(kn,n)) + ic_{e,k-1}e^{-i\phi} \sin(|g_{n}|\tau_k \xi(kn,n))=0.
\end{align}
The solution explicitly provides us with $g_{n}\tau_k.$ At this point, we have inverted the $k$th $n$JC operation. We must now invert the $k$th qubit driving step. To do that, we apply $\hat{C}_k^\dagger$ to Eq.~\eqref{eq:InvertedEq},
\begin{align}
  \hat{C}_k^{\dagger}\hat{Q}_{k}^{(n)\dagger}\ket{\Psi_{k}}=\sum_{l=0}^{k-1}\bigg[(\widetilde{c}_{g,l}\mu_{k}+i\widetilde{c}_{e,l}\nu_{k})\ket{g,ln}+(i\widetilde{c}_{g,l}\nu_{k}^*+\widetilde{c}_{e,l}\mu_{k}^*)\ket{e,ln}\bigg],   
\end{align}
where $\mu_{k}=\cos(|\Omega|\tau_k')$ and $\nu_{k}=e^{-i\theta}\sin(|\Omega|\tau_k').$
We, once again, must ensure that the highest Fock state is linked only to the qubit ground state. Thus, we must set the amplitude of $\ket{e,(k-1)n}$ to zero. This yields the constraint equation
\begin{align}\label{eq:DriveConst}
    i\widetilde{c}_{g,k-1}e^{i\theta}\sin(|\Omega|\tau_k')+\widetilde{c}_{e,k-1}\cos(|\Omega|\tau_k')=0,
\end{align}
which when solved provides us with $\Omega_k\tau^\prime_k.$ We have now inverted the $k$th qubit driving operation, and have arrived at the $(k-1)$th intermediate state, $\ket{\Psi_{k-1}}.$ Repeatedly applying the outlined inversion steps from $k=M$ to the initial state, resets the system back to $\ket{g,0}$ and provides us with the sets of parameters $\{g_{n,j}\tau\}_{j=1}^M$ and $\{\Omega_{j}\tau\}_{j=1}^M$.
\begin{figure*}[t]
\includegraphics[scale=1.05,trim={1.3cm 0cm 0 0}]{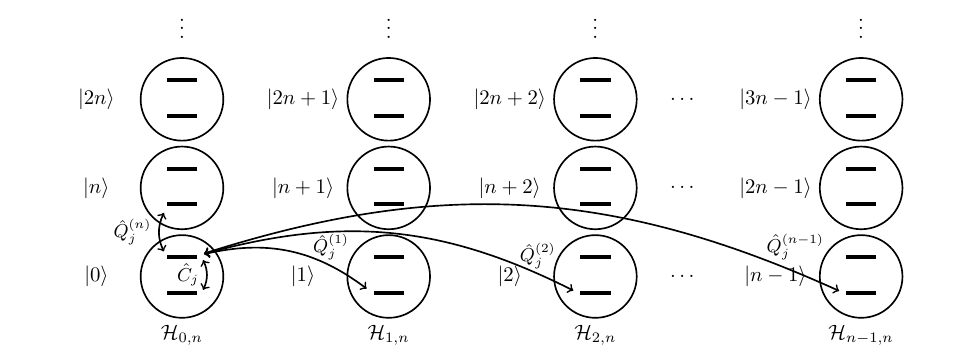}% Here is how to import EPS art
\caption{\label{fig:ControlSchematic} {Multiphoton control schematic. The vertical columns represent the subspaces $\mathcal{H}_{k,n}$, and each circle represents a Fock state with two bars; the lower and upper bars represent the qubit ground and excited states, respectively. The $n$JC operation, $\hat{Q}_{j}^{(n)}$ combined with qubit rotations, $\hat{C}_{j}$, enable arbitrary state synthesis within one column. Other interaction orders different than $n$ can be used to transition and create superpositions between states in different subspaces (columns).}} 
\end{figure*}

We represent the first step of the basic state synthesis protocol in $\mathcal{H}_{0,n}$ pictorially in the first column of Fig.~\ref{fig:ControlSchematic}. The operation of $\hat{C}_j$ induces the transitions $\ket{g,jn}\leftrightarrow\ket{e,jn}$, and $\hat{Q}_j^{(n)}$ induces transitions $\ket{e,jn}\leftrightarrow\ket{g,(j+1)n}.$ Note that if the initial state in the protocol is $\ket{g,k}$ with $k=1,2,...,n-1$, our protocol allows us to generate an arbitrary state in the respective subspace $\mathcal{H}_{k,n}.$ The protocol can be generalized to synthesize an arbitrary state in the Hilbert space with some modifications; this is discussed later in this paper.
\subsubsection{Time scaling of $n$-photon interactions}
\begin{figure}[t]
\centering
\includegraphics[scale=1,trim={.5cm .5cm 0cm 0cm}]{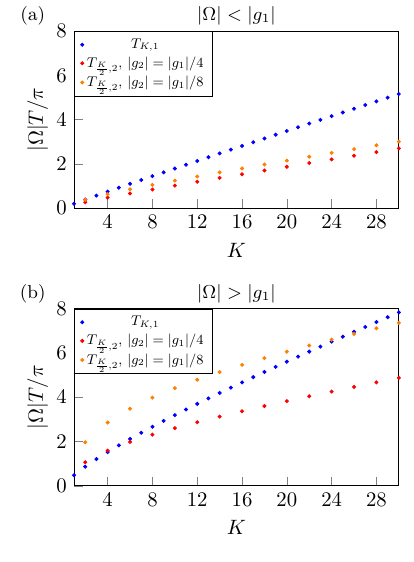}% Here is how to import EPS art
\caption{\label{fig:PrepTime} {Rotationally symmetric state synthesis time upper bound for different interaction orders as a function of number of steps. The time scaling of the upper bound in Eq.~\eqref{eq:MultiphotonTimeEstimate} is examined in different parameter regimes. In $(a)$, $|\Omega|=2\pi\times25\text{ MHz}$, and in $(b)$, $|\Omega|=2\pi\times200\text{ MHz}$. The blue dots represent $n=1$ with $|g_1| = 2\pi\times100\text{ MHz}$. In the case of $n=2$, the red dots have $|g_2|=|g_1|/4$, while the $|g_2|=|g_1|/8$ for the orange dots. The two-photon state preparation outperforms its one-photon counterpart for larger circuit depths (number of steps) in both regimes ($|\Omega|<|g_1|$ and $|\Omega|>|g_1|$). The two-photon time estimates have support exclusively on even steps since for each two steps using a linear interaction, we need one step using a two-photon interaction.}} 
\end{figure}

With the $n$LE protocol details outlined, we now examine scaling for $n$-photon interactions. Generally, we can approximate the total time to prepare a state in $K$ steps (2$K$ operations) using $n$-photon interactions as \cite{LawEberely_ArbControl_1996,Strauch_ArbControlTwoRes_2010}
\begin{align}\label{eq:MultiphotonTimeEstimate}
    T_{K,n}=\frac{K\pi}{|\Omega|}+\sum_{j=1}^K\frac{\pi}{|g_n|\sqrt{\frac{(jn)!}{((j-1)n)!}}}.
\end{align}
Here, we assume that the couplings $g_n$ and $\Omega$ can be switched on and off, but remain fixed at a constant value whenever the corresponding interaction is active. The estimate in Eq.~\eqref{eq:MultiphotonTimeEstimate} assumes that each step spans a single oscillation period, and thus provides an upper bound on the total state synthesis time even though most solutions lie below this limit. An important point to keep in mind is that $n$-photon interactions' Rabi swaps, are enhanced by the combinatorial factor $\xi(a,b)$. However, in practice, the coupling strength for a given interaction order, $n$, is typically much smaller than the coupling strength one order lower, $n-1.$ Thus, to get a good estimate, we must provide realistic coupling strengths. Here, we assume $|g_1|=2\pi\times100\text{MHz}$ and $|g_2|=2\pi\times25\text{MHz}$, which are within the range of experimentally realizable parameters \footnote{We discuss these parameter choices within the context of a realistic superconducting circuits implementation later in this paper.}\cite{Campagne-Ibarcq_QECGrid_CondDisp_2020,Eriksson_UnivControl_Cubic_2024}. For completeness, we also consider $|g_2|=2\pi\times12.5\text{ MHz}$ to explore the time scaling for different ratios of $|g_1|/|g_2|.$ We now explore two regimes: one where $|\Omega|<|g_1|$ and the other has $|\Omega|>|g_1|.$ With respect to the linear interaction, the former regime has the qubit drive as the bottleneck since it is the slowest operation, while the latter regime has the 1JC as the bottleneck. Figure~\ref{fig:PrepTime} shows the time scaling with number of steps for $|\Omega|<|g_1|$ and $|\Omega|>|g_1|,$ where we plot $ T_{K,1}^{\ket{\psi}}$ and $ T_{K/2,2}^{\ket{\psi}}.$ In the case of $|\Omega|<|g_1|$, we use $|\Omega|=2\pi\times25\text{ MHz}$. Here, the two-photon interaction always performs better for both values of $|g_2|$. For $|\Omega| > |g_1|$, we use $|\Omega|=2\pi\times200\text{ MHz}.$ In this case, we find that for a small number of steps, the linear interaction is better, and for a larger number of steps, eventually the two-photon interaction outperforms its one-photon counterpart. In general, the speed-up provided by the multiphoton interactions can be explained by the reduction in the number of steps along with the scaling of the multiphoton Rabi swap rate for the $j$th swap, $\sqrt{(jn)!/((j-1)n)!}$. We note that in realistic bosonic error correction experiments, where the auxiliary qubit is used for the control and readout of the logical oscillator mode, strong qubit drives are undesired since they can lead to spurious driving-induced transitions \cite{Sank_MIST_2016,Sivak_RealTimeQEC_2023,Dai_DUSTSpectroscopy_2025}.

Here, we only study the state preparation using the linear and two-photon interactions, and we discuss the use of interaction orders beyond $n=2$ and their time scaling in App.~\ref{app:HigherOrderControl}.
\subsubsection{Examples}
For the examples below, we work in the $|\Omega|<|g_1|$ regime and fix $|\Omega|=2\pi\times25\text{ MHz},$ $|g_1|=2\pi\times 100\text{ MHz}$ and $|g_2|=2\pi\times 25 \text{ MHz}$. The first example is a two-component even cat state,
\begin{align}
    \ket{\mathcal{C}^{+,2}_{\alpha}}=\frac{1}{\mathcal{N}}(\ket{\alpha}+\ket{-\alpha}),
\end{align}
which is in $\mathcal{H}_{0,2}$. We set $\alpha=\sqrt{2}$ and truncate our target state at $10$ photons. The linear interaction protocol requires 20 operations to obtain the two-component cat state with fidelity $\mathcal{F}=0.99999884$, while the two-photon protocol achieves the same state and fidelity requiring only 10 operations \footnote{Note that when providing fidelity estimates, we compare the prepared states in the finite state space with the untruncated target state. A cutoff for the target state is chosen to ensure a fidelity exceeding $0.999.$}. Table~\ref{tab:StatePrep} provides the list of parameters, obtained from the inversion algorithm presented earlier (solving Eqs.~\eqref{eq:JCConst} and~\eqref{eq:DriveConst}), for $n=1$ and $n=2.$ It is important to highlight that many solutions exist for the sets $\{g_{n,j}\tau_j\}_{j=1}^M$ and $\{\Omega_{j}\tau_j'\}_{j=1}^M$. Hence, we use the smallest possible solutions, which correspond to the shortest state preparation times. 

The state preparation times for $\ket{\mathcal{C}_\alpha^{+,2}}$ (with $\alpha=\sqrt{2}$) in the case of $n=1$ and $n=2$, obtained by plugging in the coupling parameters into the results from Table~\ref{tab:StatePrep}, are
$$T_{n=1}^{\ket{\mathcal{C}_\alpha^{+,2}}}= 110\text{ ns}\qquad \text{and} \qquad T_{n=2}^{\ket{\mathcal{C}_\alpha^{+,2}}}=26\text{ ns},$$respectively. In this case, the two-photon state preparation protocol provides a $76\%$ improvement over the one-photon protocol. The time estimate using Eq.~\eqref{eq:MultiphotonTimeEstimate} gives $T_{K=10,n=1}^{\ket{\mathcal{C}_\alpha^{+,2}}}\lessapprox 225\text{ ns}$ and $T_{K=5,n=2}^{\ket{\mathcal{C}_\alpha^{+,2}}}\lessapprox 128\text{ ns}$. As expected, these estimates serve as an upper bound and, in this case, the actual improvement between $n=1$ and $n=2$ is substantially better than the estimate. Generally, it is straightforward to calculate exact times for any target state using the desired coupling strengths. However, the estimate illuminates the scaling with increasing circuit depth that can quickly find the break-even depth at which the two-photon (or higher order) protocol outperforms the one-photon protocol for some parameter regime.
\begin{table*}[t]
\centering
%\resizebox{\textwidth}{!}{%
\begin{tabular}{c|cccc|cccc}
 Target states &\multicolumn{4}{c}{$\ket{\mathcal{C}_\alpha^{+,2}},\,\mathcal{F}=0.99999884$}&\multicolumn{4}{c}{$\ket{\mathcal{C}_\alpha^{++,4}},\,\mathcal{F}=0.99999744$}\\ \hline
 %&\multicolumn{4}{c}{Interaction order}&\multicolumn{4}{c}{Interaction order}\\
 Int. orders&\multicolumn{2}{c}{$n=1$}&\multicolumn{2}{c}{$n=2$}&\multicolumn{2}{c}{$n=1$}&\multicolumn{2}{c}{$n=2$}\\ \hline
 Total time& \multicolumn{2}{c}{$T=110$\text{ ns}}& \multicolumn{2}{c}{$T=26$\text{ ns}}& \multicolumn{2}{c}{$T=89$\text{ ns}}& \multicolumn{2}{c}{$T=46$\text{ ns}}\\\hline
 Step&$|g_{1}\tau_j|$&$|\Omega\tau_j'|$&$|g_{2}\tau_j|$&$|\Omega\tau_j'|$&$|g_{1}\tau_j|$&$|\Omega\tau_j'|$&$|g_{2}\tau_j|$&$|\Omega\tau_j'|$\\ \hline
 1& 1.2460 & 1.5708 & 0.8510 & 0.7397& 1.3074& 1.5708& 0.4704& 1.5708\\
 2 &0.7312 & 1.5708 & 0.3937 & 0.3290& 1.5323& 1.5708& 0.2539& 1.5708\\
 3 &0.3605 & 1.5708 & 0.1915 & 0.2926& 0.4022& 1.5708&0.0237 & 1.5708\\
 4 &0.4279 & 1.5708 & 0.0938 & 0.5195& 0.3166& 1.5708& 0.2099& 1.5708\\
 5 &0.3270 & 1.5708 & 0.1656 & 0.5745& 0.8831& 1.5708& -& -\\
 6 &0.1752 & 1.5708 &-&-& 0.0909 & 1.5708&-& -\\
 7 &1.0519 & 1.5708 & -&-& 0.5937 & 1.5708& -& -\\
 8 &0.9979 & 1.5708 & -&-&0.5554 & 1.5708& -& -\\
 9 &0.7450 & 1.5708 & - & -& -& -& -& -\\
 10& 0.4967& 1.5708 & - & -& -& -& -& -\\
 
\end{tabular}
%}
\caption{\label{tab:StatePrep}State preparation parameters for one- and two-photon Law-Eberly protocols. The states listed are two- and four-component cat states with ${\alpha}=\sqrt{2}.$ The parameters are obtained by solving Eqs.~\eqref{eq:JCConst} and~\eqref{eq:DriveConst}.} 
\end{table*}

We consider another state of importance to bosonic codes, a four-component cat state,
\begin{align}
    \ket{\mathcal{C}_\alpha^{++,4}}=\frac{1}{\mathcal{N}}(\ket{\alpha}+\ket{i\alpha}+\ket{-\alpha}+\ket{-i\alpha}),
\end{align}
with Fock support on both $\mathcal{H}_{0,2}$ and $\mathcal{H}_{0,4}$, i.e., we can use a one-, two- or four-photon interaction. We set $\alpha=\sqrt{2}$ and truncate our target state at 8 photons to obtain a fidelity of $\mathcal{F}=0.99999744$ in 8 steps (16 operations) and 4 steps (8 operations) using the one- and two-photon interactions, respectively. The exact preparation times, as presented in Table~\ref{tab:StatePrep}, are $$T_{n=1}^{\ket{\mathcal{C}_\alpha^{++,4}}}=89\text{ ns}\qquad \text{and} \quad T_{n=2}^{\ket{\mathcal{C}_\alpha^{++,4}}}=46\text{ ns},$$ where the two-photon protocol achieves a $48$\% improvement over its one-photon counterpart.

We move on to the preparation of binomial codewords \begin{align}
    \ket{0_{\text{bin}}^{(N,K)}}=\frac{1}{\sqrt{2^N}}\sum_{\text{even} \,p \in\{0,...,N+1\}}\sqrt{\begin{pmatrix}
        N+1\\p
    \end{pmatrix}}\ket{p(K+1)},
\end{align} where $N$ is the order of the code, which defines the maximum Fock support, and $K$ defines the spacing between Fock states \cite{Michael_BinomialCode_2016}. Note that these states have finite Fock support by design and thus, can be, in principle, prepared with unit fidelity using LE and $n$LE protocols. The first state we consider is $    \ket{0_{\text{bin}}^{(1,1)}}=(\ket{0}+\ket{4})/\sqrt{2},$ which belongs to both $\mathcal{H}_{0,2}$ and $\mathcal{H}_{0,4}$ like the four-component cat state. We find the exact preparation times (for $\mathcal{F}=1$) to be $$ T^{\ket{0_{\text{bin}}^{(1,1)}}}_{n=1}= 47\,\text{ns} \qquad \text{and}\qquad T^{\ket{0_{\text{bin}}^{(1,1)}}}_{n=2}= 26\,\text{ns},$$ with an improvement of $45\%$ for the two-photon interaction. We also find the preparation times for $\ket{0_{\text{bin}}^{(2,2)}}=(\ket{0}+\sqrt{3}\ket{6})/2;$ $$ T^{\ket{0_{\text{bin}}^{(2,2)}}}_{n=1}= 70\,\text{ns} \qquad \text{and}\qquad T^{\ket{0_{\text{bin}}^{(2,2)}}}_{n=2}= 38\,\text{ns},$$ with an improvement of $46\%$ for $n=2$.

We now study the preparation of GKP states using our protocol. We consider the finite-energy approximate logical zero state
\begin{align}\label{eq:GKPStates}
    \ket{0_{\text{GKP}}^{(\kappa,r,P)}}=\frac{1}{\mathcal{N}}\left(\sum_{k=-P}^P e^{- \frac{\pi \kappa^2 (k \sqrt{2 \pi})^2}{\sqrt{2 \pi}}}  \hat{D}(k\sqrt{2\pi})\hat{S}(r)\ket{0} \right),
\end{align}
where $\hat{S}(r)=\exp(r(\aop^{\dagger 2}-\aop^2)/2)$ is the squeezing operator with squeezing parameter $r$, $2P+1$ is the number of displaced squeezed states, and $\kappa$ is the width of the Gaussian envelope \cite{Gottesman_EncQubOsc_2001}. An important pair of metrics for characterizing the quality of GKP states is the effective squeezing in the $x$ and $p$ quadratures \cite{Kasper_SensorState_2017},
    \begin{align}
        \Delta_x=\sqrt{\frac{1}{2\pi}\ln(1/|\langle\hat{D}(i\sqrt{2\pi})\rangle|^2)}\qquad \text{and}
\qquad        \Delta_p=\sqrt{\frac{1}{2\pi}\ln(1/|\langle\hat{D}(\sqrt{2\pi})\rangle|^2)},
    \end{align}
where the effective squeezing is typically quoted in dB via $\Delta^{(\text{dB})}_{x/p}=-10\log_{10}(\Delta_{x/p}^2).$ Typical effective squeezing requirements for fault-tolerant quantum computing with GKP codes concatenated with an outer error correcting code are above 10~dB \cite{Noh_Surface-GKP_2020,Noh_LowOverheadSurfaceGKP_2022}. We note, though, that stabilizer expectations (subsequently, effective squeezings) do not necessarily bound fidelities \cite{Goldberg_StabFidBound_2025}  but are still a good metric for the complexity of state preparation \cite{Terhal_EncodQubitPhaseEstim_2016,Hastrup_MeasFreeGKPPrep_2021}. 

 We consider the set of parameters $(\kappa,r,P)=(0.15,1,2)$ which define a GKP state with $\Delta_x^{(\text{dB})}=8.69$~dB and $\Delta_p^{(\text{dB})}=13.38$~dB. We truncate the state at 36 photons and obtain a fidelity of $\mathcal{F}=0.99930617$. This requires 36 steps (72 operations) and 18 steps (36 operations) using the one- and two-photon interactions, respectively. We find the exact preparation times to be
$$ T^{\ket{0_{\text{GKP}}^{(0.15,1,2)}}}_{n=1}= 384\,\text{ns} \qquad \text{and}\qquad T^{\ket{0_{\text{GKP}}^{(0.15,1,2)}}}_{n=2}= 200\,\text{ns},$$
which shows a $48\%$ improvement using the two-photon interaction. Finally, we consider a higher-quality GKP states with parameters $(\kappa,r,P)=(0.15,1.5,3)$ with $\Delta_x^{(\text{dB})}=13.03$~dB and $\Delta_p^{(\text{dB})}=15.17$~dB. The state is truncated at 76 photons and achieves a fidelity of $\mathcal{F}=0.99939574$. State synthesis requires 76 and 38 steps with the one- and two-photon interactions, respectively. The time estimates for these states are the longest considered thus far with
$$ T^{\ket{0_{\text{GKP}}^{(0.15,1.5,3)}}}_{n=1}=797 \,\text{ns}\qquad \text{and} \qquad T^{\ket{0_{\text{GKP}}^{(0.15,1.5,3)}}}_{n=2}= 388 \,\text{ns},$$
where  the two-photon protocol provides an improvement of $52\%.$

The parameters depicted in Table~\ref{tab:StatePrep} were obtained by solving the constraint equations numerically at each step in the inversion and selecting the smallest values of $|g_{n}\tau_k|$ and $|\Omega\tau_k'|$. These values correspond to only one of the many possible solutions that satisfy the constraint equations. We note that in Table~\ref{tab:StatePrep}, for many of the qubit drive operations $|\Omega \tau_k'| = \pi/2\approx1.5708$. These specific qubit driving strength values occur due to the structure of the target state and the order of the employed $n$JC interaction used in the generation protocol. For example, consider the inversion algorithm using a $n=2$ JC interaction for a four-component cat state, which has Fock state populations in multiples of four, i.e., $\ket{C_\alpha^{++,4}} = \sum_{j}c_j|4j\rangle$. As a result of the four-photon symmetry and the 2JC interaction used, all $|4j\rangle$ Fock states share the same qubit state, either $|e/g\rangle$, throughout the inversion. Simultaneously, the Fock states $|4j +2\rangle$ are all coupled to the orthogonal qubit state. This will mathematically yield $\widetilde{c}_{g,k-1} = 0$ in Eq.~\eqref{eq:DriveConst} for each step in the inversion, reducing the constraint equation to the simplified form, $\cos(|\Omega_k|\tau) = 0,$ which produces the $\pi/2$ qubit drive solutions. All states where qubit populations alternate between $\ket{e/g}$ at $n$-photon spacings when using a $n$JC interaction for the inversion, will have $\pi/2$ solutions for the qubit drive. In Table~\ref{tab:StatePrep}, we see that the only state without $\pi/2$ qubit drive solutions is the two-component cat state when synthesized using a $n=2$ JC interaction. As this cat state has populations at two-photon spaced Fock states, using a two-photon interaction will create superpositions of $\ket{e/g}$ for a given Fock state during the inversion. Therefore, the solutions to Eq.~\eqref{eq:DriveConst} become non-trivial and no longer simply $\pi/2$. For the two-component cat state, we see that alternating qubit states at two-photon multiples are not possible due to the spacing of the Fock states relative to the JC interaction order used.

We further present state preparation times for all of the above states with all combinations of a stronger qubit drive, $|\Omega|=2\pi\times200 \text{MHz},$ as well as a weaker two-photon coupling, $|g_2|=2\pi\times 12.5\text{MHz}$, in Fig.~\ref{fig:MethodComp}. The results are consistent with the time scaling arguments presented earlier and shown in Fig.~\ref{fig:PrepTime}.

%To date, the 2JC interaction has not been realized experimentally in superconducting circuits. However, third order nonlinear interactions (same order as the 2JC) have been realized \cite{Sandbo_ThreePhotonSPDC_2020,Eriksson_UnivControl_Cubic_2024}, and coupling a qubit to these realizations is not too difficult. Thus, the parameters we assume for the 2JC are within a reasonable range of what has been achieved. With the examples we will present, we will make a strong case for the utility of realizing such a two-photon interaction in near-term experiments.   

\subsection{Numerically optimized general sequences}\label{sec:NumericalSeq}

For the 1JC interaction, there are known schemes that go beyond the LE protocol. One approach is to simply numerically optimize the operation sequence for a target oscillator state with the final qubit state being in the ground state \cite{Liu_HybridQOProc_2024}. Additionally, the JC interaction and qubit drive could be operated simultaneously and allow for the qubit-oscillator dressed state transition to be driven, which provides a speed up in some cases compared to the LE protocol \cite{Strauch_AllResControl_2012}.

We now introduce a more general setup where we have a sequence, exactly as in the $n$LE protocol, without the intermediate steps' constraints, i.e., we allow the qubit and oscillator to be arbitrarily entangled throughout the intermediate steps. We only require that the system start in $\ket{g}\ket{0}$ and end in $\ket{g}\ket{\Psi_t}.$ Explicitly, we want to prepare the target unitary of depth $M$\begin{align}
\hat{U}_t^{n\text{JC}}(\vec{\vartheta})=\prod_{j=1}^{M}\hat{Q}^{(n)}_{j}\hat{C}_j,
\end{align} with $\vec{\vartheta}=(\Omega \tau_1',g_n\tau_1,...,\Omega\tau_M',g_n\tau_M)$ parameterizing the $n$JC operators and qubit rotations. Given a target state and a depth $M$ sequence of operations, we want to maximize the fidelity,\begin{align}
    \mathcal{F}_M(\vec{\vartheta})=|\bra{g}\bra{\Psi_t} P_N \hat{U}_t^{n\text{JC}}(\vec{\vartheta})\ket{g}\ket{0}|^2,
\end{align} where $P_N=\sum_{l=0}^N \dyad{l}$.  Generally, the fidelity is a monotonically increasing function of $M$. We solve this optimization problem numerically and find a solution $\vec{\vartheta}$ with a fidelity associated with it. We initialize the unitary with random numbers and then optimize them for many iterations. We use tens of random seeds per calculation and for each seed, we use between 5000-10000 iterations and allow for tolerances up to $10^{-9}.$ We rely on \texttt{scipy.optimize} \cite{Virtanen_SciPyPaper_2020} for all of our calculations. Then, we evaluate the state preparation times for all possible combinations of qubit drive and two-photon coupling strengths used earlier. We also perform the same calculations and provide state preparation times for the linear interaction.

\begin{figure}[t]
\centering
\includegraphics[scale=.935,trim={0cm 0cm 0cm 0cm}]{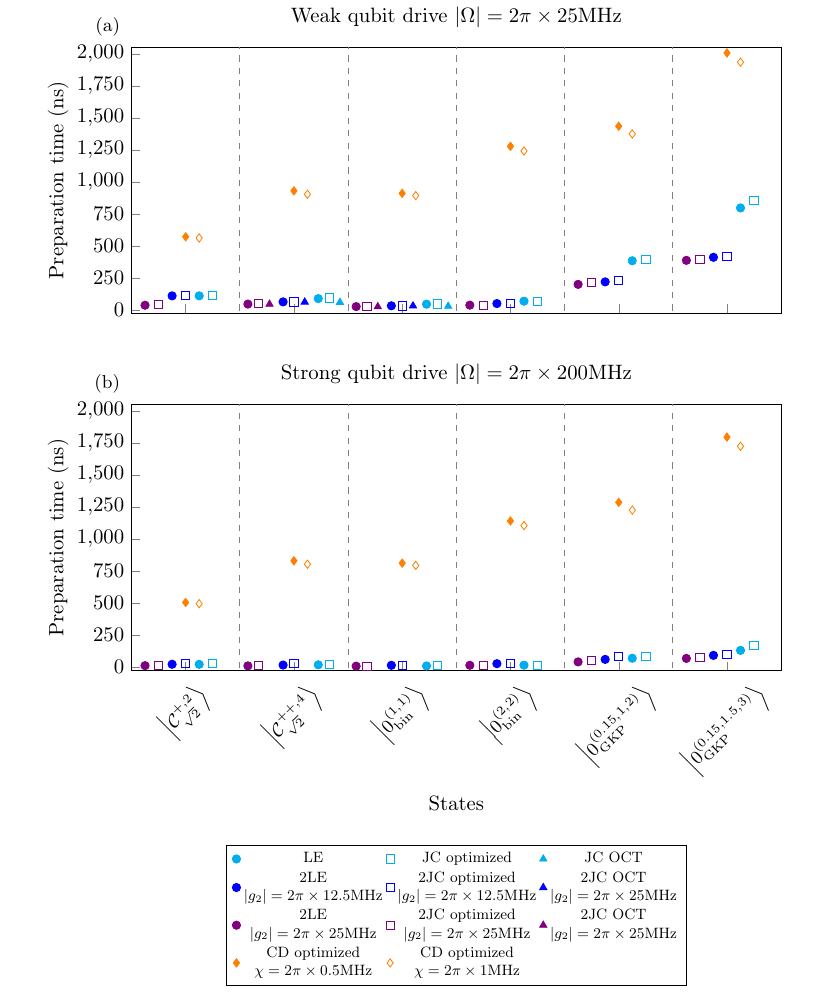}% Here is how to import EPS art
\caption{\label{fig:MethodComp} {Time estimates for preparing bosonic codewords using various control schemes.}} 
\end{figure}

The optimized calculations should, in principle, perform as well or better than the $n$LE protocol. We find for the cases we consider, the numerical optimization schemes very closely recover the results of the $n$LE protocol or perform slightly worse. We present the time estimates for the previously discussed bosonic codeword examples in Fig.~\ref{fig:MethodComp}. These results suggest that for the specific circuit structure of alternating $n$JC and qubit drive operations, the $n$LE protocol is optimal. Other studies on the linear interaction that improve upon the LE protocol employ the JC interaction and qubit drive simultaneously, where they additionally employ multi-tone driving on the qubit \cite{Strauch_AllResControl_2012}.
\subsection{Comparison to conditional displacement control}
We briefly described conditional displacement control earlier in this paper. Conditional displacements have been experimentally demonstrated and used for bosonic codeword state preparation in three-dimensional superconducting microwave cavities \cite{Eickbusch_FastUnivCont_Disp2022,Sivak_RealTimeQEC_2023, Brock_QuditGKPQEC_2025}. We now provide a detailed comparison of using sequences of conditional displacements and qubit rotations versus the $n$LE protocol for preparing the example states discussed above. First, we outline the procedure from which we get the parameters for a given circuit. Then, we provide time estimate for each state using realistic parameters suitable for this control scheme.

Similarly to the numerically optimized $n$JC sequences, we take the system to start in $\ket{g}\ket{0}$ and end in $\ket{g}\ket{\Psi_t}.$ Explicitly, we want to prepare the target unitary composed of a sequence of depth $M$ operations\begin{align}
\hat{U}_t^{\text{CD}}(\vec{\vartheta})=\prod_{j=1}^{M}\hat{U}_{\text{CD},j}\hat{C}_j,
\end{align} with $\vec{\vartheta}=(\Omega \tau_1',\beta_1,...,\Omega\tau_M',\beta_M)$ parameterizing the qubit rotations and conditional displacements. Here, $\beta_k$ is the $k$th displacement parameter. We discuss the gate time of a conditional displacement below. For a target state and a depth $M$ sequence of operations, we want to maximize the fidelity,\begin{align}
    \mathcal{F}_M(\vec{\vartheta})=|\bra{g}\bra{\Psi_t} P_N \hat{U}_t^{\text{CD}}(\vec{\vartheta})\ket{g}\ket{0}|^2.
\end{align} The calculations are a carried out as described in the previous section with the CD unitary replacing the $n$JC unitary.

Before we provide comparisons for each state, we should discuss the time estimate for implementing a conditional displacement gate. The conditional displacement gate can be implemented by driving the oscillator and auxiliary qubit in a particular sequence with the qubit-oscillator system situated in the dispersive regime; this particular gate implementation is known as echoed-conditional displacement (ECD) \cite{Eickbusch_FastUnivCont_Disp2022}. Alternatively, the conditional displacement gates can be implemented by driving qubit near or at cross-resonance with the oscillator \cite{Touzard_GateConditionalDisp_2019,Ayyash_ResSchCats_2024}. Here, we assume the former implementation---ECDs. Following Ref.~\cite{Eickbusch_FastUnivCont_Disp2022}, the ECD gate implementation sequence consists of two qubit rotations, four displacements, and a waiting period to allow for the accumulation of the desired conditional phase between pulses. For the $j$th ECD gate, the waiting period $T_j$ is determined by $\beta_j = 2i\alpha\sin(\chi T_j/2)$, where $\beta_j$ is the desired conditional displacement amplitude, $\chi$ is the dispersive shift of the qubit-oscillator system, and $|\alpha| = \alpha_0$ sets the maximum displacement amplitude allowed by experimental constraints. Thus, the total state preparation time using the ECD protocol is given by the sum of the qubit rotations, displacements and waiting periods,
\begin{equation}
    T_{\text{ECD}} = \sum_{j=1}^{M}\bigg(\tau_j^\prime + \frac{\pi}{2\Omega}+T_j + \frac{\alpha_0}{\epsilon} + 2\frac{\alpha_0\cos(\chi T_j/4)}{\epsilon} + \frac{\alpha_0\cos(\chi T_j/2)}{\epsilon}\bigg).
\end{equation}
Here, $\epsilon$ is the displacement drive strength. For all calculations presented below, we set $\epsilon = 2\pi\times200\text{ MHz}$, which is within the range of realized oscillator drive strengths \cite{Eickbusch_FastUnivCont_Disp2022}. Furthermore, we present time estimates for various parameter regimes. For this control scheme, we have additional parameters that enter into the calculation: the oscillator drive strength $\epsilon$ and the dispersive shift $\chi=|g_1|^2/|\Delta|.$ For the dispersive shift, we parameterize it by assuming a fixed $|g_1|=2\pi\times100\text{ MHz}, $ and thus, the variable we can adjust is the qubit-oscillator detuning, $\Delta.$ We provide estimates using $\chi=2\pi\times 0.5\text{MHz}$ and $2\pi\times1\text{MHz}$ that are typical of a planar superconducting circuit in the dispersive regime as well as weak and strong qubit drives with $|\Omega|=2\pi\times 25 \text{MHz}$ and $|\Omega|=2\pi\times200\text{MHz}$.

The time estimates of the optimized CD calculations for the same bosonic codewords as before are presented in Fig.~\ref{fig:MethodComp}. The times presented are the shortest times for circuit depths reaching fidelity greater than $0.999$ except in the case of the GKP states; GKP states are known to require very large depths to reach high fidelities \cite{Eickbusch_FastUnivCont_Disp2022} and as such, we settle for circuit depths achieving GKP state  fidelities greater than 0.99. We note that for all states prepared using LE/$n$LE, a threshold fidelity of 0.999 is used. Remarkably, we find the state preparation times of LE, two-photon LE and their numerical variants to be substantially shorter than those required by the CD control scheme. Intuitively, this is expected as the linear and $n$-photon Rabi swaps become faster for higher energy levels, where as the ECD gates do not depend on any energy level structure and are pretty much only dependent on the displacement size at a given step. Additionally, this is not exclusive to the ECD implementation of conditional displacements since also qubit-driven conditional displacements \cite{Touzard_GateConditionalDisp_2019,Ayyash_ResSchCats_2024} are only dependent on the displacement of a given step.

In a more recent work \cite{Singh_NAQSP_2025}, some analytic prescriptions were given for preparing various bosonic states such as squeezed states and GKP states using sequences of conditional displacements and qubit rotations rather than numerically optimized sequences as above. Of the states examined in our paper, Ref.~\cite{Singh_NAQSP_2025} gave analytic sequences for preparing two-component cat states and multi-peak GKP states as described in Eq.~\eqref{eq:GKPStates}. We decomposed the sequences for two-component cat and GKP states into ECDs and qubit rotations and found their time estimates (using the same $\epsilon$ and combinations of $\chi$ and $\Omega$) to perform worse than the numerically optimized conditional displacement sequences. We briefly discuss one such sequence used to prepare GKP states. Consider the finite-energy GKP state \begin{align}
    \ket{1_{\text{GKP}}^{(\kappa,r,P)}}=\frac{1}{\mathcal{N}}\left(\sum^P_{k=-P} e^{- \pi \kappa^2 ((k+1/2) \sqrt{2 \pi})^2/\sqrt{2 \pi}} \hat{D}(\sqrt{2 \pi} (k+1/2)) \hat{S}(r) |0\rangle\right).
\end{align} We set $\kappa=\Delta/(2\pi)^{1/4},~r = -\log(\Delta)$ and $P=3$ with $\Delta=0.34$. Then, Ref.~\cite{Singh_NAQSP_2025} shows that \begin{align}
    \ket{1_{\text{GKP}}^{(\Delta/(2\pi)^{1/4},-\log(\Delta),3)}}\simeq \prod_{j=0}^2 e^{i\frac{\pi\Delta^2}{4(j+1)\alpha}\hat{\sigma}_x \hat{p}}e^{i\frac{\pi}{4(j+1)\alpha}\hat{\sigma}_y \hat{x}} e^{2i\alpha \hat{\sigma}_x \hat{p}}\hat{S}(r)\ket{0}.
\end{align}
Then, using another sequence in Ref.~\cite{Singh_NAQSP_2025} to prepare the initial squeezed state, we find the total time (squeezing plus subsequent operations) to prepare this state under the most optimistic ECD assumptions ($\epsilon=2\pi\times200\text{MHz},\,\chi=2\pi\times1\text{MHz},\, |\Omega|=2\pi\times200 \text{MHz}$) to be about $3.3\,\mu\text{s},$ whereas preparing this state with the most pessimistic two-photon parameters ($|g_2|=2\pi\times12.5\text{MHz},\,|\Omega|=2\pi\times 25\text{MHz}$) using 2LE takes about $232\,\text{ns}.$ The work of Ref.~\cite{Singh_NAQSP_2025} is the result of concatenating non-commuting conditional displacements with quantum signal processing composite pulses. It is interesting to consider the concatenation of quantum signal processing composite pulses such as BB1 \cite{Wimperis_BB1_1994} with sideband control sequences. We leave that as a future direction.

We note that these significant improvements of sideband-based controls over conditional displacement control come with practical limitations.  The inherent advantage of the sideband-based controls (linear and nonlinear) is that the gate time is shorter at larger circuit depths because the higher energy levels possess faster swap times. This advantage past some energy level, depending on the specifics of the system and experimental setup, turns into a limitation as it becomes impractical to rapidly "switch on and off" the sideband interaction to alternate with the qubit drive. We discuss the practical limitations in greater detail in Sec.~6. For the cat and binomial states presented, the substantial improvements are not limited by the practical limitations since they are truncated at pretty low Fock states, whereas the high cutoffs used for GKP states definitely cross over into the regime where the shortness of gate time becomes impractical.

\subsection{Optimal control theory at the Hamiltonian level}

We previously found  in Sec.~\ref{sec:NumericalSeq} that for a sequence of alternating $n$JC and qubit drive operations, the $n$LE protocol is optimal. We now investigate a more general approach. The state preparation time can often be reduced by using
numerically optimized pulses, instead of the step-by-step state
preparation protocol. We perform optimal control theory (OCT)
calculations using the gradient-ascent pulse engineering (GRAPE) algorithm \cite{Khaneja_GRAPE_2005} to obtain such pulses and determine the minimum time needed
to prepare various states. Since the main purpose of these
calculations is to demonstrate the possibility and conditions for
speeding up the state preparation process using optimized pulses, we
consider a situation in which the qubit-oscillator coupling term is fixed,
while control fields can be applied to all three qubit operators
($\sigX$, $\sigY$, and $\sigZ$) without any constraints. This
situation therefore corresponds to the one limited by the qubit-oscillator
coupling and as such, we quote preparation times in terms of the coupling strength parameter. This somewhat simplified model captures the main physical
mechanisms at play while avoiding unnecessarily complicating the
control problem with constraints. The calculations generally follow
those described in Ref.~\cite{Basyildiz_SpeedLimitQutritGates_2025}. We include Fock states up to 20 photons in the simulations. We
divide the total pulse time into $10^3$ steps. We initialize the
control pulses at each time step to small random values in the range
$[ -0.2 \pi g_n, 0.2\pi g_n]$. We apply $10^5$ pulse optimization
iterations. We vary the pulse duration and identify the minimum state
preparation time as the time needed to achieve fidelity
$\mathcal{F}=0.999$.

For the binomial state $(\ket{0}+\ket{4})/\sqrt{2}$, using numerically
optimized pulses, we find a state preparation time of $0.99/g_2$,
which is only slightly shorter than the $n$LE state preparation time
($1.01/g_2$). The small difference can be attributed to the fact that
we set the threshold $\mathcal{F}=0.999$ to identify the minimum state
preparation time. In other words, OCT produces the same state preparation time as the $n$LE protocol. This result is not surprising if we consider the
dynamics in the $n$LE protocol: first, a qubit $\pi$ pulse transfers the
full population from $\ket{g}\ket{0}$ to $\ket{e}\ket{0}$. Then, the $n$JC
interaction transfers half of the population to $\ket{g}\ket{2}$. Then a
second qubit $\pi$ pulse transfers the populations of $\ket{e}\ket{0}$ and
$\ket{g}\ket{2}$ to $\ket{g}\ket{0}$ and $\ket{e}\ket{2}$, respectively. Finally, the
nJC interaction transfers the full population of $\ket{e}\ket{2}$ to
$\ket{g}\ket{4}$. Intuitively, this sequence should be the fastest way to
prepare the target state, $(\ket{0}+\ket{4})/\sqrt{2}$. Our OCT results support this conclusion. If
we use single-photon JC interactions to prepare the same state, we
obtain a time of $3.41/g_1$ when using numerically optimized pulses,
which is again only slightly shorter than the time obtained using the
single-photon LE protocol ($3.45/g_1$). This result is quite surprising, because it is not obvious why we
should not obtain any speedup by using OCT. The population
dynamics that corresponds to the numerically optimized pulse follows
the dynamics of the $n$LE protocol, which indicates that the OCT
algorithm is finding the $n$LE pulse sequence. We cannot exclude the
possibility that there are better pulses than the ones that were found
using our numerical algorithm. However, based on the previous example
and our past experience with similar OCT problems \cite{Ashhab_SpeedLimitsMultiqubit_2012,Basyildiz_SpeedLimitQutritGates_2025},
we expect that our results are accurate to within about 10\%. We
therefore do not expect much speedup from using OCT-optimized pulses in
this case.

{For the four-component cat state $\ket{\mathcal{C}_{\sqrt{2}}^{++,4}}$ studied earlier with a Fock cutoff of 8, we obtain a state
preparation time of $0.94/g_2$ with optimized pulses, which is
slightly shorter than the time obtained from the $n$LE protocol
($0.96/g_2$). Using single-photon JC interactions, we obtain a time of
$3.81/g_1$ when using optimized pulses. This number is roughly 35\%
shorter than the $n$LE time of $5.68/g_1$. There is no simple
explanation for why this case is the only one that gave a speedup (and
a quite significant speedup at that) from using optimized pulses.}

{The OCT results for the binomial and cat state are plotted in Fig.~\ref{fig:MethodComp}(a) and contextualized with the LE/$n$LE and conditional displacement preparation times.}

{Our OCT calculations reveal another interesting and rather surprising
result: The state $(\ket{0}+\ket{2})/\sqrt{2}$ can be prepared faster
than the $n$LE state preparation time, even though the $n$LE protocol
involves only one qubit rotation followed by one JC interaction. The
optimal time ($0.95/g_2$) obtained using optimized pulses is about
15\% shorter than the $n$LE time ($1.11/g_2$). This result is
reminiscent of another surprising result in which two-qubit gates can
be accelerated by temporarily exciting the qubits outside the
computational space to ancillary high energy states to utilize the
strong interactions in those states \cite{Ashhab_SpeedLimitsMultiqubit_2012}. However, in this system, reaching higher energy
levels requires going through the state $\ket{g}\ket{2}$ first, making this
mechanism unsuitable to explain the speedup in this case. The more
intuitive explanation is that, in the $n$LE protocol, we transfer half
the population from $\ket{g}\ket{0}$ to $\ket{e}\ket{0}$ and then transfer the
full population of $\ket{e}\ket{0}$ to $\ket{g}\ket{2}$ in a time given by
$\pi/(2^{3/2}g_2)$. In contrast, when using optimized pulses, we can
initially drive the occupation probability of the state $\ket{e}\ket{0}$
above 0.5, which leads to a faster transfer of population from
$\ket{e}\ket{0}$ to $\ket{g}\ket{2}$ compared to the $n$LE protocol. Indeed, our
numerical results for the optimal-time pulse show that the population
of the state $\ket{e}\ket{0}$ quickly rises to 0.99 in the early stages of
the dynamics, before gradually decreasing to 0 at the end of the
control pulse. Of course this approach requires careful control of the
dynamics to bring part of the population back from $\ket{e}\ket{0}$ to
$\ket{g}\ket{0}$ while avoiding leakage of $\ket{g}\ket{2}$ to $\ket{e}\ket{2}$ and
higher energy levels. Nevertheless, a speedup compared to the $n$LE
protocol can be achieved. A similar calculation for preparing the
state $(\ket{0}+\ket{1})/\sqrt{2}$ with single-photon JC interactions
gives a state preparation time of $1.48/g_1$ when using optimized
pulses, which is about 6\% shorter than the LE time of $1.57/g_1$.}

\section{Arbitrary State Preparation}\label{sec:ArbControl}

In this section, we introduce a protocol that can synthesize arbitrary states in the Hilbert space using a combination of interaction orders and (photon-number) selective qubit rotations to introduce transitions between different $\mathcal{H}_{k,n}$ subspaces, as shown in Fig.~\ref{fig:ControlSchematic}. We explain the selective qubit rotations and how they can be implemented using any interaction order in the dispersive regime (linear or multiphoton). Then, we describe our protocol for arbitrary state synthesis, which we call \textit{fine-tune-then-populate} for reasons that will become apparent later. Next, we elucidate the relationship between $n$JC interactions and other control schemes. Finally, we discuss the connection between our control protocols and dissipative stabilization.

\subsection{Combination of different interaction orders for arbitrary control}

In a naive attempt to generalize the multiphoton protocol across all $\mathcal{H}_{k,n}$ subspaces, we consider combining different interaction orders as follows. One could first employ a linear interaction (LE scheme) to populate the lowest energy states across all $\mathcal{H}_{k,n}$ subspaces (bottom of the columns in Fig.~\ref{fig:ControlSchematic}). Next, one could apply qubit drives and $n$JC interactions to populate the required Fock states within each $\mathcal{H}_{k,n}$, thereby generating the target state. We will now show that, in its current form, this approach does not provide arbitrary control over all subspaces. As an example, let $n=2$ and consider that the state $(\ket{0} + \ket{1})/\sqrt{2}$, which has populations in both $\mathcal{H}_{0,2}$ and $\mathcal{H}_{1,2}$, that have already been generated. After a subsequent qubit drive, we can now perform the following two-photon swap to start populating higher levels in the $\mathcal{H}_{0,2}$ and $\mathcal{H}_{1,2}$ subspaces:
\begin{equation}
    \ket{e}\frac{(\ket{0} + \ket{1})}{\sqrt{2}} \rightarrow \ket{g}\frac{(\ket{2} + \ket{3})}{\sqrt{2}}.
\end{equation}
Applying $\hat{Q}^{(2)}$ to this initial state yields
\begin{align}
    \hat{Q}^{(2)}~\ket{e}\frac{(\ket{0} + \ket{1})}{\sqrt{2}} =  \frac{1}{\sqrt{2}} \bigg[& \cos(|g_2| \tau \sqrt{2}) \ket{e,0} + \cos(|g_2|\tau \sqrt{6}) \ket{e,1} \nonumber \nonumber\\&- ie^{-i\phi}\bigg(\sin(|g_2|\tau \sqrt{2}) \ket{g,2} + \sin(|g_2|\tau \sqrt{6}) \ket{g,3}\bigg)\bigg].
\end{align}
For simplicity, consider only the value of $|g_2|\tau$ that guarantees that applying $\hat{Q}^{(2)}$ completes a full swap for each Fock state. This would require the constraint equations, 
$$
    \cos(|g_{2}|\tau \sqrt{2}) =0\qquad  \text{and}\qquad
    \cos(|g_{2}|\tau \sqrt{6}) =  0,     
$$ 
to be solved simultaneously. However, no analytical solution for this exists, as the ratios of the two cosine arguments are incommensurable. Generally, a perfect simultaneous swap of both states could occur only if the arguments inside the cosine functions are integer multiples of $\pi/2$. In our example, we show that it is generally not possible for complete excitation swaps for multiple Fock states. The resulting state after applying $\hat{Q}^{(2)}$ would have combinations of $\ket{e/g}$ qubit states in at least one of the Fock states. When applying further pairs of qubit drives and $n$JC interactions, spurious excitations attributed to incomplete swaps for some Fock state leads to contradictions, like presented above, when solving for coupling parameters. Simply combining different orders of interactions cause spurious excitations as the Fock states are no longer spaced in $n$-photon multiples, making solving the constraint equations generally intractable. Thus, to have sufficient control over the different $\mathcal{H}_{k,n}$ subspaces, we must be able to individually address Fock states, which can be achieved through selective qubit rotations. Selective qubit rotations enable us to swap only certain Fock states to higher excitations while leaving the others in the ground state, thus controlling which columns in Fig.~\ref{fig:ControlSchematic} are populated. By modifying our naive protocol to include selective qubit rotations, we show in the following sections that arbitrary control is possible and outperforms LE when the state is sparsely populated. 

\subsubsection{Selective qubit rotations}

In the dispersive regime of a linear qubit-oscillator interaction, i.e., $g_1\ll|\omega_q-\omega_o|$, the resulting effective interaction, $\sigZ\adag\aop$, can be used to achieve selective qubit rotations. To rotate the qubit conditionally on the oscillator state $\ket{l}$, we set the qubit driving frequency to \cite{Strauch_ArbControlTwoRes_2010} 
$\omega_d=\omega_q+\chi^{(1)}(1+2l),$
where $\chi^{(1)}=g_1^2/(\omega_q-\omega_o)$ is the linear dispersive shift. This effectively realizes the unitary
\begin{align}\label{eq:SelectiveRotation}
    \hat{C}_j^{\ket{l}}=\dyad{l}\hat{C}_{j} + \sum_{k\neq l}\dyad{k}.
\end{align}

{In the case of multiphoton interactions, the dispersive regime for an interaction of order $n$ is defined by $g_n\ll|\Delta_n|,$ where $\Delta_n=\omega_q-n\omega_o$ and $\chi^{(n)}=g_n^2/\Delta_n$. Here, the interaction Hamiltonian reads (to second order in $g_n/\Delta_n$)~\cite{Ayyash_MultiphotonDispersive_2025}
\begin{align}
    \hat{H}^{(n)}_{\text{Disp}}=\frac{\chi_n}{2}\sum_{k=0}^n C_{n,k}^{+}\sigZ(\adag\aop)^k+\frac{\chi_n}{2}\sum_{k=1}^{n-1} C_{n,k}^{-}(\adag\aop)^k,
\end{align}
where $C_{n,k}^+=(-1)^{n+k}s_1(n+1,k+1)+s_1(n,k)$ with $s_1(a,b)$ being Stirling numbers of the first kind.} Similarly to the linear case, we can achieve a qubit rotation conditioned on Fock state $\ket{l}$ using $n$-photon interaction in the dispersive regime by adding a qubit drive and setting the drive frequency to $\omega_d=\omega_{l}^{(n)}$ with
\begin{align}
    \omega_{l}^{(n)}=\omega_q+\chi^{(n)}\sum_{k=0}^{n}C_{n,k}^+(l)^k.
\end{align}
This process also leads to the unitary in Eq.~\eqref{eq:SelectiveRotation}. The qubit driving strength used in the case of selective rotations must obey $\Omega <|\chi^{(n)}|$, regardless of the interaction order employed, to remain within the dispersive perturbative description's regime of validity \cite{Strauch_ArbControlTwoRes_2010,Ayyash_MultiphotonDispersive_2025}. Additionally, we want to avoid strong qubit drives that can result in non-resonant transitions. Thus, we can generate selective rotations using any qubit-oscillator interaction order in its dispersive regime in the presence of a qubit drive.

\subsubsection{Fine-tune-then-populate}

\begin{figure}[t]
\centering
\includegraphics[scale=.9,trim={0cm 0cm 0cm 0cm}]{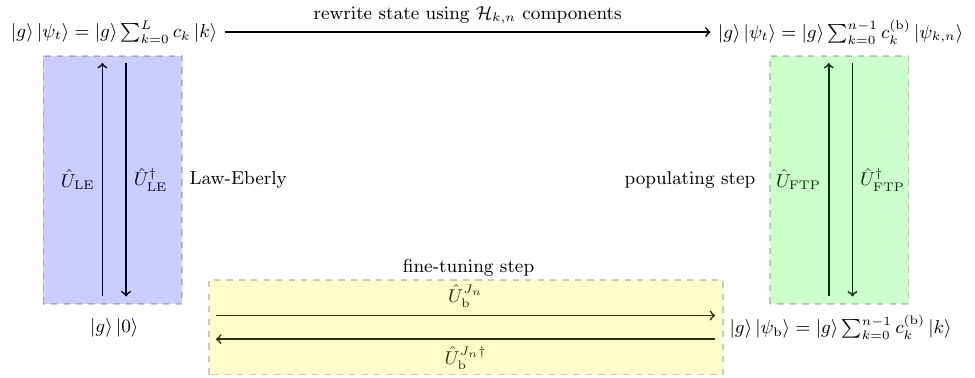}% Here is how to import EPS art
\caption{\label{fig:FTPSchematic} {Fine-tune-then-populate (FTP) protocol schematic. The blue box symbolizes the Law-Eberly protocol to prepare an arbitrary finite Fock cutoff state. FTP is composed of a fine-tuning step, shown in the yellow box, that prepares a base state. Once the base state is prepared, each component of the base state is evolved independently in the populating step, shown in the green box, where individual $\mathcal{H}_{k,n}$ subspaces (the columns in Fig.~\ref{fig:ControlSchematic}) are populated.}} 
\end{figure}

\begin{figure}[t]
\centering
\includegraphics[scale=1,trim={.5cm .5cm 0cm 0cm}]{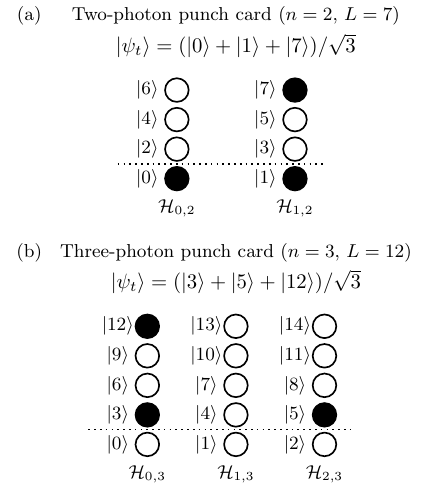}% Here is how to import EPS art
\caption{\label{fig:PunchCard} {Arbitrary state preparation punch card. For a given target state, a shaded circle indicates a nonzero Fock state contribution, while an unshaded circle means zero contribution. The number of steps required to prepare a target state is equal to the height in each column, where the height is determined by the largest shaded Fock state above the dotted line, and the additional steps used to generate the base state. The entries of the height vector, $\vec{h}^{(n)}=(h_0^{(n)},h_1^{(n)},...,h_{n-1}^{(n)})$, correspond to the height of each column. For example, the state in (a) has $\vec{h}^{(n)}=(0,3)$, and the state in (b) has $\vec{h}^{(n)}=(4,0,1).$ }} 
\end{figure}
With the selective qubit rotations described, we may now proceed to describe an algorithm that uses a combination of multiphoton interactions to synthesize an arbitrary state. {The protocol consists of two phases, which give it its name: a \textit{fine-tune} phase, in which we prepare a base state whose amplitudes are the overlap of the target state with the $\mathcal{H}_{k,n}$ subspace projectors (see details below), and a \textit{populate} phase, in which we climb each symmetry subspace to fill in the remaining Fock states.} Let the target state be some general state $\ket{\psi_t}=\sum_{k=0}^Lc_{k}\ket{k},$ which could have support over all $\mathcal{H}_{k,n}$ subspaces. We must first fix $n$ to be the highest-order interaction we use. {We begin by populating the oscillator with some superposition of the lowest states in each of the $\mathcal{H}_{k,n}$ subspaces, i.e., preparing the \textit{base} state $\ket{\psi_\text{b}}=\sum_{k=0}^{n-1}c_k^{(b)}\ket{k}$ with $c_k^{(b)}=\bra{\psi_t}P_{k,n}\ket{\psi_t},$ where $P_{k,n}=\sum_{l=0}^\infty \dyad{ln+k}$ . Explicitly, these coefficients appear in the target state as $\ket{\psi_t}=\sum_{l=0}^{n-1}c_{l}^{(\text{b})}\ket{\psi_{k,n}}$, where $\ket{\psi_{k,n}}\in\mathcal{H}_{k,n}$ and $\ket{\psi_{k,n}}=P_{k,n}\ket{\psi_t}/c_{k}^{(b)}$. } We prepare this superposition state using $J_n$ steps, where we refer to a step as a pair of operations involving a qubit drive (selective or otherwise) followed by some $n$JC interaction, namely, each step involves 2 operations. {Using the LE protocol, the base state requires $J_n=n-1$ steps, since the $n$ consecutive Fock states $\ket{0},\ket{1},\dots,\ket{n-1}$ are populated one photon at a time. For small $n$, LE is the most efficient path (in number of steps) to generating the base state; for larger $n$, the count can be reduced by populating the base state with higher-order interactions in bundles rather than single photons.\footnote{{For $n=2$, a qubit drive followed by a 1JC operation gives $J_n=1$. For $n=3$, two such sequences give $J_n=2$. For $n=6$, two repetitions of a qubit drive followed by a 2JC operation, then a qubit drive followed by a 1JC operation, give $J_n=3$---fewer than the $n-1=5$ steps the LE protocol would require. In practice, because of the smaller coupling strengths at higher orders and the presence of spurious interactions, the highest order we can realistically employ is $n=2$ or $n=3$, requiring 1 or 2 steps, respectively.}}} We refer to the unitary that prepares the base state as $\hat{U}_{\text{(b)}}^{J_n}$ with $J_n$ denoting the number of steps it uses. 

Once the base state is generated, we can exclusively use selective qubit rotations along with the specified (largest) $n$JC to efficiently populate the oscillator with all other Fock states. {Concretely, each column is climbed one rung at a time: a selective rotation $\hat{C}^{\ket{jn+k}}$ excites the qubit only when the oscillator occupies the specific Fock state $\ket{jn+k}$, after which the $n$JC interaction swaps this excitation upward by $n$ photons, $\ket{e,jn+k}\to\ket{g,(j+1)n+k}$. Because the rotation addresses a single Fock state, the incommensurate-swap obstruction discussed above never arises and the constraint equations remain solvable.} The largest Fock states, $\ket{L}$, can be expressed as $L= jn+k$ for some $k\in\{0,1,...,n-1\}.$ Then, the fine-tune-then-populate (FTP) unitary sequence that takes the base state to the target state is 

\begin{align}
    \hat{U}_{\text{FTP}}=\prod_{j=1}^{\lfloor L/n\rfloor}\prod_{k=0}^{n-1}\left(\hat{Q}_{j,k}^{(n)}\hat{C}_{j,k}^{\ket{jn+k}}\right),
\end{align}
where index $j$ represents the row crossing each subspace in Fig.~\ref{fig:ControlSchematic}, index $k$ represents the column (the particular $\mathcal{H}_{k,n}$ subspace), and the selective rotations are conditioned on the Fock state in the $j$th row and $k$th column, $\ket{jn+k}.$ {Each factor $\hat{Q}_{j,k}^{(n)}\hat{C}_{j,k}^{\ket{jn+k}}$ constitutes one step (one rung) of the populate phase.} Here, $\lfloor L/n \rfloor$ signifies the row which the largest Fock state--in the target state--occupies ($\ket{L}$). Then, the total number of steps for arbitrary control is at most
\begin{align}
    K_{\text{arb}}(n,L)=J_n+L - (n-1).
\end{align}
Note that it is always the case that $L>n$. This upper bound on the number steps exactly matches LE when $J_n=n-1.$ The upper bound improves when $J_n-(n-1)<0.$ For $n=2$ and $n=3$, $K_{\text{arb}}(2,L)=K_{\text{arb}}(3,L)=L$, while for $n=6$, $K_{\text{arb}}(6,L)=L-2$---a tighter bound than LE.

With this upper bound in mind, we take an illustrative example to show how the structure of the state significantly affects the number of steps actually needed. We introduce a \textit{`punch card'} description for the state synthesis that simplifies counting the number of steps involved. {In this description, the \textit{height} $h_k^{(n)}$ of a column is the rung of its highest occupied Fock state, counted upward from the base row (the dotted line); equivalently, it is the number of $n$JC swaps needed to reach the top of that column.} Figure~\ref{fig:PunchCard} shows the state preparation punch card for two different states relying on two different highest-order interactions. A shaded circle in the punch card indicates a non-zero Fock coefficient, while an unshaded circle indicates a zero coefficient of the respective Fock state. In Fig~\ref{fig:PunchCard}(a), we consider the example state $(\ket{0}+\ket{1}+\ket{7})/\sqrt{3},$ where the highest-interaction order is two and the largest Fock state is $\ket{7}$. The number of steps required to generate this state is 1+3 with 1 step coming from the base state and 3 from the rest of the protocol. This is calculated by going column-by-column on the punch card and adding the height (in number of circles above the dotted line) of the largest shaded Fock state. Carrying out the same counting procedure in Fig.~\ref{fig:PunchCard}(b) we find the number of steps to be 2+5 with 2 steps from generating the base state and 5 steps from the rest of the protocol. Both these examples are well below the upper bound provided above. Generally, the total number of steps for the preparation of arbitrary states can be calculated as
\begin{align}
    N_{\text{arb}}(n,\vec{h}^{(n)})= J_n + \sum_{k=0}^{n-1}h_{k}^{(n)},
\end{align}
where $\vec{h}^{(n)}=(h_{0}^{(n)},\dots,h_{n-1}^{(n)})$ and $h_{k}^{(n)}$ is the height of the $\mathcal{H}_{k,n}$ column in the punch card of the target state.

This suggests that the exact structure of the state and distribution of its Fock support over the subspaces significantly affects the number of steps needed to generate the state. {In particular, the FTP step count is governed by the \textit{sum of the per-column maximum heights}, $\sum_k h_k^{(n)}$, whereas the LE count is governed by the \textit{single global} maximum Fock index $L$. The FTP protocol is therefore advantageous when the support is spread thinly across subspaces, so that no individual column climbs high---even if some Fock state reaches a large absolute index. Increasing $n$ distributes the support over more columns and lowers each column's height, which is the main lever the protocol provides.} The question of state preparation time is nontrivial and depends on $n$, $L$, and the strength of the interactions used along with the structure of the Fock coefficients. We now provide a time estimate for a given number of steps. Let $T_{\text{b}}$ and $T_{\text{c}}^{(n)}$ be the time estimates for preparing the base state and fine-tuning then populating the columns (in Fig.~\ref{fig:ControlSchematic}), respectively. Then, the time estimate for using the FTP protocol is
\begin{align}\label{eq:FTPTime}
    T_{{L,n}}^\text{FTP}=T_{\text{b}}+T_{\text{c}}^{(n)},
\end{align}
where
\begin{align}
    T_{\text{c}}^{(n)}=\sum_{k=0}^{n-1} \left(h_{k}^{(n)}\frac{\pi}{\Omega}+\sum_{j=1}^{h_{k}^{(n)}}\frac{\pi}{g_n\sqrt{\frac{(jn+k)!}{((j-1)n+k)!}}}\right).
\end{align}
This estimate assumes a fixed coupling $g_n$ and $\Omega$ throughout the state preparation procedure. Here, the driving strength must obey $\Omega<|\chi^{(n)}|.$ {This constraint is the hidden cost of the populate phase: the selective rotations must be weak to remain in the dispersive regime, so each populate step is correspondingly slow. It is this per-step slowdown---traded against the reduced number of steps---that ultimately determines whether FTP outperforms LE.} For simplicity, we assume all the base states are prepared using a linear interaction via LE. The time estimate for preparing a state with maximum Fock support ending at $\ket{L}$ using LE is
\begin{align}
    T^{\text{LE}}_L=\sum_{j=1}^{L}\frac{\pi}{g_1\sqrt{j}};
\end{align}
the base state preparation time is obtained by setting $L=n-1.$
We now compare the time estimate of our FTP protocol with that of LE. For the case of LE, we assume $g_1=2\pi\times100\text{ MHz}$ and $|\Omega|=2\pi\times25\text{ MHz}.$ For FTP, we set $n=2$ and assume $g_1=2\pi\times100\text{ MHz}$, $g_2=2\pi\times 25\text{ MHz}$, and $|\Omega|=2\pi\times25\text{ MHz}.$ As a first example, we consider the state $$\ket{\psi_t}=(\ket{0}+\ket{2}+\ket{5}+\ket{9})/\sqrt{4}.$$ The time estimate for preparing this state using LE is
$$T^{\text{LE}}_{L=9}\lessapprox 200\text{ ns},$$
whereas the the time estimate for using FTP is
$$ T^{\text{FTP}}_{L=9,n=2}\lessapprox 181\text{ ns}.$$ Here, we used our punch card description to find $\vec{h}^{(2)}=(1,4)$ and plugged it into Eq.~\eqref{eq:FTPTime}.
In this case, FTP yields a faster state preparation time. This example is representative of a state that is lacking $n$-fold rotational symmetry (not exclusively in $\mathcal{H}_{k,n}$ for any $k$) and is sparsely populated thus gaining a speed up from our FTP protocol. We now consider a second example state, $$\ket{\psi_t}=\sum_{k=0}^{10}\ket{k}/\sqrt{11}.$$
Preparing this state via LE takes
$$T^{\text{LE}}_{L=10}\lessapprox 222\text{ ns},$$
while using FTP requires
$$T^{\text{FTP}}_{L=10,n=2}\lessapprox275\text{ ns},$$
which was found using $\vec{h}^{(2)}=(5,4).$ In this case, FTP performs, as anticipated, worse than LE since the state is not sparsely populated. 

Generally, there are no exact conditions under which FTP performs better than LE. The sparsity discussed here is qualitative and difficult to quantify. {Quantitatively, the relevant notion is that of \textit{low maximum height per column}: the FTP protocol wins when $J_n+\sum_k h_k^{(n)}$, weighted by the per-step times above, totals less than the LE protocol time.} Thus, to determine whether FTP is a better protocol for a given state, one must use the punch-card description (as we did above) to calculate the time estimate for each state, which is a simple and fast calculation.
\subsection{Connection to other control schemes}

We previously discussed the phase-space instruction set using conditional displacement operations combined with single-qubit rotations. The multiphoton generalization of a conditional displacement is the conditional $n$-photon squeezing gate, defined as
\begin{align}
    \hat{U}_{\text{CS}_{n}}(\zeta)=\dyad{g}\hat{S}_n(\zeta)+\dyad{e}\hat{S}_n(-\zeta),
\end{align}
where $\hat{S}_n(\zeta)=\exp(\zeta\adagn-\zeta^*\aop^n)$ is the $n$-photon squeezing operator \cite{Braunstein_GenSqz_1987, Ashhab_GenSqz_2025}\footnote{Note that these generalized squeezing operators are ill-defined as their generators are not essentially self-adjoint for $n>2$, and they require further regulating terms to ensure self-adjointness \cite{Fischer_SelfAdjSqz_2025,Ashhab_FiniteDimSqz_2025}. However, we use them here symbolically to mean the realization of $n$-photon squeezing operations in the presence of further regulating terms that are negligible on short timescales. These operations have been experimentally realized in different platforms such as superconducting circuits \cite{Sandbo_ThreePhotonSPDC_2020,Eriksson_UnivControl_Cubic_2024} and trapped ions \cite{Saner_GenArbSuperpositionSqz_2024}.}. In a similar manner to the improvements our use of the $n$JC interaction provides to the sideband instruction set, higher-order conditional $n$-photon squeezing gates enable more efficient control in the phase-space instruction set \cite{Ayyash_DrivenTwoPhoton_2024,Liu_HybridQOProc_2024}. The conditional displacement and 1JC operations are, up to a first-order Trotter decomposition, related by \cite{Liu_HybridQOProc_2024}
\begin{align}
    \hat{U}_{\text{CD}}(gt)=&\text{H}\hat{R}_x(\pi)\hat{Q}^{(1)}(gt)\hat{R}_x(\pi)\hat{Q}^{(1)}(gt)\text{H}+ O((gt)^2),
\end{align}
where $\text{H}$ is the qubit Hadamard gate and $\hat{R}_x(\pi)=\exp(-i\pi\sigX).$ This last equation allows for a seamless translation between the phase-space and sideband instruction sets. This is also the case for the $n$JC and conditional $n$-photon squeezing operations,
\begin{align}
    \hat{U}_{\text{CS}_n}(gt)=&\text{H}\hat{R}_x(\pi)\hat{Q}^{(n)}(gt)\hat{R}_x(\pi)\hat{Q}^{(n)}(gt)\text{H}+O((gt)^2).
\end{align}
Some applications such as bosonic phase estimation are more naturally implemented using conditional displacement and squeezing gates \cite{Terhal_EncodQubitPhaseEstim_2016, Ayyash_DrivenTwoPhoton_2024}. Thus, it is useful to translate between instruction sets depending on the intended use.

\subsection{Concatenation with dissipative stabilization}

Some bosonic codes benefit from dissipative stabilization schemes, where the codewords are steady states of a particular open system, e.g., two- and four-component cat states \cite{Mirrahimi_DynProt_CatQubits_2014, Gilles_TwoPhCat_1993} and GKP states \cite{Royer_GKPStab_2020,Nathan_GKPDissip_2025,Sellem_GKPDissip_2025}. We focus on the cases of two- and four-component cat states whose stabilizing dissipators are those enacting two- and four-photon losses on the oscillator.

The multi-component cat state stabilization schemes rely on a Kerr-nonlinear oscillator Hamiltonian with a parametric $n$-photon drive on the oscillator, $\hat{H}_n=\omega_o\adag\aop + G_n\adagn\aop^n + \widetilde{\Omega}_n(t)(\adagn +\aop^n),$ along with an effective $n$-photon dissipation enacted on the oscillator. Explicitly, the effective open system master equation for a state stabilized by an $n$-photon dissipator is \cite{Mirrahimi_DynProt_CatQubits_2014,Gautier_CombinedConfinement_2022}
\begin{align}
   \frac{d}{dt}\hat{\rho}=-i[\hat{H}_n,\hat{\rho}]+\mathcal{D}(\aop^n)\hat{\rho} ,
\end{align}
where $\mathcal{D}(\hat{O})\hat{\rho}=\hat{O}\hat{\rho}\hat{O}^\dagger - \{\hat{O}^\dagger\hat{O},\hat{\rho}\}/2.$ First, we note that the physical mechanisms in superconducting circuits (see Sec.~\ref{sec:cQED}) giving rise to the $n$JC interaction inadvertently result in additional $n$-photon self-Kerr, $\adagn\aop^n,$ and $n$-photon parametric driving, $\adagn+\aop^n,$ terms that can be tuned to realized $\hat{H}_n$. Second, the requisite $n$JC interactions needed for our specific and arbitrary control protocols can serve a dual purpose and be concatenated with dissipative stabilization. For example, we can use the specific two-photon control protocol to prepare the two-component cat state after which the two-photon qubit-oscillator interaction can remain switched on. This interaction, $\sigP\aop^2 + \sigM\adagT$, can be seen from the perspective of the oscillator, as enacting two-photon dissipation with the coupling strength effectively tuning the two-photon loss rate \cite{Mirrahimi_DynProt_CatQubits_2014,Leghtas_TwoPhotonDissip_CatExp_2015}. This is also the case for the four-photon JC interaction that can be used to prepare a four-component cat state and then stabilize it by acting as a four-photon dissipator \cite{Mirrahimi_DynProt_CatQubits_2014,Vaneslow_FourPhotonDissip_2025}.

In the case of a two-component cat state, this serves as an alternative initialization protocol to adiabatically \cite{Gilles_TwoPhCat_1993,Gilles_TwoPhCat2_1993,Gerry_TwoPhCats_1993} or counterdiabatically \cite{Puri_TwoPhotonCat_2017} transition from $\ket{0}$ and $\ket{1}$ to the even- or odd-parity two-component cat states, respectively.

\section{Generalization to Multiple Oscillators}\label{sec:Multimode}

\begin{figure}[t]
\centering
\includegraphics[scale=.9]{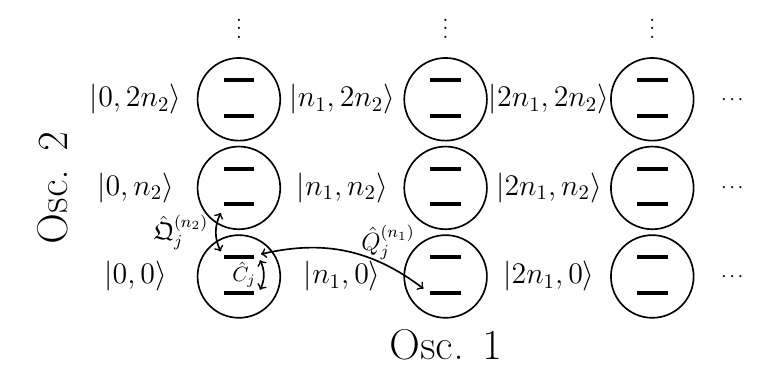}% Here is how to import EPS art
\caption{\label{fig:multimode} {Schematic for multiphoton control over the two-oscillator subspace $\mathcal{H}_{0,n_1}\otimes\mathcal{H}_{0,n_2}.$ Each circle represents a tensor product of Fock states in $\mathcal{H}_{0,n_1}\otimes\mathcal{H}_{0,n_2}.$ The lower and upper horizontal bars represent the qubit ground and excited states, respectively. The operators $\hat{{Q}}^{(n_1)}_{j}$ and $\hat{\mathfrak{Q}}^{(n_2)}_{j}$ represent $n_1$JC and $n_2$JC operations on oscillator 1 and 2, respectively.}} 
\end{figure}
The single-mode protocols described in Secs.~\ref{sec:Multi_control} and~\ref{sec:ArbControl} can be generalized to multiple oscillators. In particular, we generalize the protocol of Ref.~\cite{Strauch_ArbControlTwoRes_2010}, which employs linear interactions and selective qubit rotations, to achieve multiphoton control within the $\mathcal{H}_{0,n_1}\otimes\mathcal{H}_{0,n_2}$ subspace. Then, we extend our arbitrary state preparation protocol to the case of multiple oscillators with multiphoton interactions.

Extending upon the single-oscillator setup, we consider a system with one qubit and two oscillators, where each oscillator is coupled to the qubit via a multiphoton JC interaction. The $n$JC interactions have the additional freedom that allows for each oscillator to be of a possibly different order ($n_1$ and $n_2$). In the interaction picture, the qubit's interaction with the $k$th oscillator near $n_k$-photon resonance reads

\begin{align}
    \hat{H}^{I,n_k}=&g_{n_{k}} (t)\sigP\aop^{n_k}_k+g^{*}_{n_k}(t)\sigM a^{\dagger n_k}_k, 
\end{align}
where $g_{n_{k}}(t)$ is the coupling strength of the $n_k$-photon interaction for the $k$th oscillator. The time-evolution of oscillator 1 (2) generated by the $n_1$JC ($n_2$JC) interaction is represented by $\hat{Q}^{(n_1)}\,~(\hat{\mathfrak{Q}}^{(n_2)})$. We further require selective qubit rotations for the multi-oscillator case. The selective qubit rotations required for individual control over a state $\ket{l_1, l_2}$ can be generalized from Eq.~\eqref{eq:SelectiveRotation} for two oscillators 
\begin{align}\label{eq:SelectiveRotationMutliOscillator}
    \hat{C}_j^{\ket{l_1,l_2}}=\dyad{l_1,l_2}\hat{C}_{j} + \sum_{\substack{k_1\neq l_1\\ k_2\neq l_2}}\dyad{k_1, k_2},
\end{align}
where we drive the qubit in the dispersive regime at the frequency,
\begin{align}\label{eq:MultiOscillatorDispersiveShift}
    \omega_{l_1,l_2}^{(n_1,n_2)} =& \omega_q + \chi^{(n_1)}_1 \sum^{n_1}_{k=0} C^+_{n_1, k} ~ (l_1)^k + \chi^{(n_2)}_2 \sum^{n_2}_{k=0} C^+_{n_2, k} ~ (l_2)^k. 
\end{align}
Here, $\chi^{(n_k)}_k$ denotes the $n_k$-photon dispersive shift of the $k$th oscillator, while $l_1$ ($l_2$) are the selectively driven Fock states that are multiples of $n_1$ ($n_2$). 

Consider the target state $$\ket{\Psi_t}=\sum_{k_1,k_2=0}^{L_1,L_2}c_{k_1,k_2}\ket{k_1n_1,k_2n_2},$$ where the largest populated Fock state is $\ket{L_1 n_1,L_2n_2}$. The protocol for synthesizing the target state in the two-oscillator subspace $\mathcal{H}_{0,n_1}\otimes\mathcal{H}_{0,n_2}$ is given by

\begin{align}\label{eq:multi_target}
    \ket{\Psi_{t}}=&\hat{U}_t\ket{g,0,0} \bigg(\prod_{i=0}^{{L_2 -1}}\hat{U}_{2,i}\bigg)\prod_{k=0}^{L_1-1} \hat{Q}^{(n_1)}_{k}\hat{C}_k^{\ket{k n_1, 0}}\ket{g,0,0} 
\end{align}
with 
\begin{align}
    \hat{U}_{2,i} = \prod_{j=0}^{L_1} \hat{\mathfrak{Q}}^{(n_2)}_{i,j}\hat{C}_{i,j}^{\ket{j n_1, i n_2}}.
\end{align}
The diagram in Fig.~\ref{fig:multimode} illustrates the protocol from Eq.~\eqref{eq:multi_target} and serves as the basis for the following discussion. In the first step of the protocol, a sequence of selective qubit rotations\footnote{For the sequence of qubit rotations used in populating the first oscillator, the qubit rotations need not be conditional. We keep them that way for notational consistency and clarity.} and $n_1$JC interactions with the first oscillator lead to populated Fock states $\ket{0,0}, \ket{n_1,0}, \ket{2n_1,0},\ldots, \ket{L_1n_1,0}$. Graphically, this is equivalent to populating the first row in Fig.~\ref{fig:multimode}, where the second oscillator remains in the vacuum state. Next, we seek to populate the Fock states in the second oscillator by employing a sequence of selective qubit rotations together with $n_2$JC interactions. Following the schematic in Fig.~\ref{fig:multimode}, we populate the row above to add excitations to the second oscillator, proceeding sequentially through the columns. For example, a selective qubit drive and $n_2$JC interaction are performed to take $\ket{0,0} \rightarrow\ket{0,n_2}$, followed by more pairs of interactions to take $\ket{n_1,0} \rightarrow\ket{n_1,n_2},~ \ket{2n_1,0} \rightarrow\ket{2n_1,n_2},~ \ldots$, until all columns in that row are completed.  This protocol is then iteratively applied to each subsequent row above until the largest desired Fock state in the second resonator, $\ket{L_2n_2}$, is reached.

Following the single mode case, an inversion can be performed on Eq.~\eqref{eq:multi_target} that allows for the coupling coefficients to be solved for by sequentially removing photons for each row and column until reaching the initial state $\ket{g,0,0}$. The constraint equations outlined in Sec.~\ref{sec:Multi_control_basic_prot} are equivalent to the equations required to solve for coupling and drive strengths for the multi-oscillator case, except we now must consider $\hat{C}^\dagger_j$ to be selective. We provide examples of state synthesis in $\mathcal{H}_{0,n_1}\otimes\mathcal{H}_{0,n_2}$ using multiphoton interactions, and we contrast the number of steps and time estimates to the state preparation relying on linear interactions in App. \ref{app:MultiOsc}. We find similar trends to those in the single oscillator case.  

Thus far, we have shown that multiphoton interactions can generate arbitrary states within $\mathcal{H}_{0,n_1}\otimes\mathcal{H}_{0,n_2}$. To synthesize any desired state by combining different orders of JC interaction for multiple oscillators, we proceed in the same manner as the single-mode case. We assume the target state to be $\ket{\Psi_t}=\sum_{k_1,k_2=0}^{L_1,L_2}c_{k_1,k_2}\ket{k_1,k_2},$ where the largest Fock state is $\ket{L_1,L_2}.$ First, we must synthesize the two-mode base state, 
\begin{align}
        \ket{\psi_b}= \sum_{k_1=0}^{n_1-1}~\sum_{k_2=0}^{n_2-1}c_{k_1,k_2}\ket{k_1,k_2}. 
\end{align}
We now generalize the fine-tune-then-populate sequence of unitaries to the case of two oscillators, which reads
\begin{align}
\hat{U}_{\text{FTP}} = \hat{U}_{\text{FTP}}^{(2)} \prod_{k_4=0}^{n_2-1}~ \prod_{k_5=0}^{n_1-1}\prod_{k_6=0}^{\lfloor L_1/n_1\rfloor-1} \hat{Q}^{(n_1)}_{k_4,k_5, k_6}  \hat{C}^{\ket{k_6 n_1 + k_5,k_4}}_{k_4,k_5, k_6},
\end{align}
with
\begin{align}
    \hat{U}_{\text{FTP}}^{(2)} =\bigg( \prod_{k_1=0}^{n_2-1}~\prod_{k_2=0}^{\lfloor L_2/n_2\rfloor-1}\prod_{k_3=0}^{L_1} \hat{\mathfrak{Q}}^{(n_2)}_{k_1,k_2,k_3} \hat{C}^{\ket{k_3, k_2  n_2 + k_1}}_{k_1,k_2,k_3} \bigg )
\end{align}
The upper bound in number of steps for this protocol can be found in an analogous manner to the single-oscillator; it reads
\begin{align}
    &K_{\text{arb}}(n_1,L_1;n_2,L_2)= J_{n_1,n_2} + n_2\bigg( L_1 - (n_1-1)\bigg)
    + (L_1+1) \bigg( L_2 - (n_2-1)\bigg).
\end{align}
where $J_{n_1,n_2}$ is the number of steps required to generate the base state. Similarly to the single-oscillator case, this upper bound corresponds to the linear multi-oscillator case in Ref.~\cite{Strauch_ArbControlTwoRes_2010}. For more precise number of steps, a multidimensional generalization of the punch card in Fig.~\ref{fig:PunchCard} can be devised. Then, the number of steps are    
\begin{align}
    &N_{\text{arb}}(n_1,n_2,\boldsymbol{h}^{(n_1,n_2)},\boldsymbol{\mathfrak{h}}^{(n_1,n_2)})=J_{n_1,n_2} + \sum_{k_1=0}^{n_1-1}\sum_{k_2=0}^{n_2-1}h_{k_1,k_2}^{(n_1,n_2)} + \sum_{k_1=0}^{L_1}\sum_{k_2=0}^{n_2 - 1}\mathfrak{h}_{k_1,k_2}^{(n_1,n_2)}.
\end{align}
Here, $\boldsymbol{h}^{(n_1,n_2)}$ is a $(n_1-1)\times(n_2-1)$ matrix whose entries $h_{k_1,k_2}^{(n_1,n_2)}$ represent the heights associated with the first oscillator in the $\mathcal{H}_{k_1,n_1}\otimes\mathcal{H}_{k_2,n_2}$ columns of the multidimensional punch card representation of the target state. The $(L_1 + 1)\times(n_2-1)$ matrix $\mathfrak{h}^{(n_1,n_2)}$ encodes the column heights of the second oscillator. In this representation, the element $\mathfrak{h}_{k_1,k_2}^{(n_1,n_2)}$ is the column height in the $\mathcal{H}_{k_2,n_2}$ subspace of the second oscillator when the first oscillator occupies Fock state $\ket{k_1}$. For instance, the value of $\mathfrak{h}_{0,0}^{(n_1,n_2)}$ would be the height of the first column in Fig.~\ref{fig:multimode}. Examples using our protocol and how it outperforms the linear case in number of steps when the state is sparsely populated are provided in App.~\ref{app:MultiOsc}.

\section{{Practical Considerations}}\label{sec:cQED}
\renewcommand{\arraystretch}{1.2}
\begin{table*}[t]

\centering
\begin{tabular}{c | c}
   
   Parameter& Two-photon operation values \\
   \hline
   $\omega_q$&$2\pi\times$\SI{10}{\giga\hertz}\\
   $\omega_o$&$2\pi\times$\SI{5}{\giga \hertz}\\
   $g_2$ & $2\pi\times$\SI{25}{\mega\hertz}\\
   $\widetilde{g}_{e1}$ & $2\pi\times$\SI{1.08}{\giga\hertz} \\
   $\widetilde{g}_{e2}$ & $2\pi\times$\SI{1.34}{\giga\hertz} \\
   $\widetilde{g}_{e3}$ & $2\pi\times$\SI{20}{\mega\hertz} \\
   $\widetilde{g}_{e4}$ & $2\pi\times$\SI{10}{\mega\hertz} \\
   $\widetilde{g}_{e5}$ & $2\pi\times$\SI{20}{\mega\hertz} \\
   $\widetilde{g}_{c}$ & $2\pi\times$\SI{30}{\mega\hertz} \\
   $\gamma_{q,r}$ & \SI{20}{\kilo\hertz}\\
   $\gamma_{o,r}$ & \SI{20}{\kilo\hertz}\\
   $\gamma_{q,\phi}$ & \SI{110}{\kilo\hertz}\\
   $\gamma_{o,\phi}$ & \SI{110}{\kilo\hertz}\\
   
\end{tabular}

\caption{\label{tab:CircuitQEDTable}Hamiltonian and decoherence parameters used in circuit QED open system simulations. The Hamiltonian parameters (excluding all $\gamma$'s) are taken from Ref.~\cite{Ayyash_DrivenTwoPhoton_2024}.}
\end{table*}
{In this section, we consider practical aspects of our proposed protocols. We begin by considering a realistic circuit QED implementation that hosts the requisite 2JC interactions. We simulate the implementation including spurious terms and decoherence effects by means of a Lindblad master equation. Since the master equation describes average effects rather than individual quantum trajectories (on a shot-to-shot basis), we further discuss mitigating oscillator and ancilla decoherence shot-to-shot by examining their effects depending on where in the state preparation circuit they occur. Finally, we discuss practical limitations in implementing multiphoton interactions and selective qubit rotations.}
\\

\subsection{Circuit QED implementation}
In our previous calculations, we relied on superconducting circuit parameters from the literature used on an idealized qubit-oscillator system. In this section, we study the state preparation of rotationally symmetric states using a realistic circuit QED system implementing 2JC interactions, which includes spurious terms as well as qubit and oscillator energy relaxation and dephasing.

There are a number of superconducting circuit proposals that can realize an effective 2JC interaction \cite{Felicetti_TwoPhotonQRM_FluxQubits_2018,Zou_TwoPhoton_ChargeQubits_2020,Ayyash_DrivenTwoPhoton_2024,Stolyarov_TwoPhotonSQUID1_2025,Stolyarov_TwoPhotonSQUID2_2025}. Here, we consider a tunable transmon coupled to an LC resonator via an asymmetric SQUID proposed in Ref.~\cite{Ayyash_DrivenTwoPhoton_2024}. When the qubit is tuned to near two-photon resonance with the oscillator, the effective system Hamiltonian with the SQUID interaction expanded to fourth order under the two-level approximation reads as \footnote{{We clarify that the implementation discussed here, based on Ref.~
\cite{Ayyash_DrivenTwoPhoton_2024}, relies tuning the qubit to two-photon resonance with the oscillator such that the 2JC interaction is the dominant interaction term. This is quite different from other implementations where the qubit and oscillator sit at particular (fixed) frequencies, and the nonlinear coupler between them is driven at specific frequencies to enable nonlinear interactions \cite{Lemonde_HardwareEfficientBosonicPerspective_2024}.}} \cite{Ayyash_DrivenTwoPhoton_2024}
\begin{align}\label{eq:cQEDHamTLA}
    \hat{H}_{\text{cQED}}=&\frac{\omega_q}{2}\sigZ + \omega_o\adag\aop - \widetilde{g}_{e4}(\adag+\aop)^3-\widetilde{g}_{e5}\sigZ(\adag+\aop)+g_2(\sigP+\sigM)(\adag+\aop)^2\nonumber\\ &-\widetilde{g}_c(\sigP-\sigM)(\adag-\aop)-(\widetilde{g}_{e1}-\widetilde{g}_{e3})(\sigP+\sigM)-(2\widetilde{g}_{e5}-\widetilde{g}_{e2})(\adag+\aop), 
\end{align}
where all couplings with a tilde ($\widetilde{g}$) correspond to spurious terms arising from the SQUID coupler. The last two terms in Eq.~\eqref{eq:cQEDHamTLA} can be eliminated as they are linear driving terms on the qubit and oscillator, which can be tuned to zero \textit{in situ} by applying cancellation drives to the qubit and oscillator. Thus, the Hamiltonian excluding the linear terms is
\begin{align}\label{eq:cQEDHamTLAFinal}
    \hat{H}_{\text{cQED}}=&\frac{\omega_q}{2}\sigZ + \omega_o\adag\aop - \widetilde{g}_{e4}(\adag+\aop)^3-\widetilde{g}_{e5}\sigZ(\adag+\aop)+g_2(\sigP+\sigM)(\adag+\aop)^2\nonumber\\ &-\widetilde{g}_c(\sigP-\sigM)(\adag-\aop).
\end{align}
In the state preparation sequence, we assume the qubit is flux-tuned to two-photon resonance with the oscillator such that the effective Hamiltonian above describes the dynamics.
When the qubit and oscillator are far detuned, we take them to be effectively decoupled.  The qubit drive is modeled by Eq.~\eqref{eq:QubitDrive}. Furthermore, we include qubit and oscillator energy relaxation and dephasing modeled by the Lindblad master equation
\begin{align}\label{eq:LindbladME}
    \frac{d}{dt}\hat{\rho}=&-i[\hat{H},\hat{\rho}]+ \gamma_{q,r}\mathcal{D}(\sigM)\hat{\rho}+ \frac{\gamma_{q,\phi}}{2}\mathcal{D}(\sigZ)\hat{\rho}+ \gamma_{o,r}\mathcal{D}(\aop)\hat{\rho} + \gamma_{o,\phi}\mathcal{D}(\adag\aop)\hat{\rho},
\end{align}
where $\gamma_{q,r}$ ($\gamma_{o,r}$) and $\gamma_{q,\phi}$ ($\gamma_{o,\phi}$) are the qubit (oscillator) relaxation and dephasing rates, respectively. Table~\ref{tab:CircuitQEDTable} lists the Hamiltonian and decoherence parameters and the values we employ in our calculations. The decoherence parameters we use are achievable within state-of-the-art planar superconducting hardware \cite{Putterman_HWEfficientBosonicConcat_2025,Eriksson_UnivControl_Cubic_2024,Hajr_KerrCat_2024}. {For a weakly anharmonic transmon, the third level, $\ket{f}$, is of concern. The $\ket{e}\leftrightarrow\ket{f}$ transition is typically detuned below the $\ket{g}\leftrightarrow\ket{e}$ transition by an anharmonicity $A\sim150-250\text{MHz}$ \cite{Putterman_HWEfficientBosonicConcat_2025}. This means the $\ket{g}\leftrightarrow \ket{f}$ transition is largely unaffected. When the qubit drive is resonant with the $\ket{g}\leftrightarrow\ket{e}$ transition and its strength is much lower than the anharmonicity, $A$, such as in our case, the amplitude of the $\ket{f}$ state can be approximately obtained as \cite{allen1975optical}
\begin{align}
    c_{\ket{f}}\simeq \frac{\sqrt{2}\Omega}{2A}c_{\ket{e}}.
\end{align}
For the qubit driving we consider here, $|\Omega|=2\pi\times25\text{MHz}$, the leakage population is between $P_{\ket{f}}=|c_{\ket{f}}|^2\sim 0.005\% -0.0139\%$. Thus, for the open system parameters we consider, the leakage to the $\ket{f}$ is negligible. For stronger qubit drives, it becomes more significant. However, as discussed before, it is more favorable to maintain weaker qubit drives to avoid spurious qubit-oscillator transitions \cite{Dai_DUSTSpectroscopy_2025}.}
\begin{figure}[t]
\centering
\includegraphics[scale=.9,trim={0cm .5cm 0cm 0.25cm}]{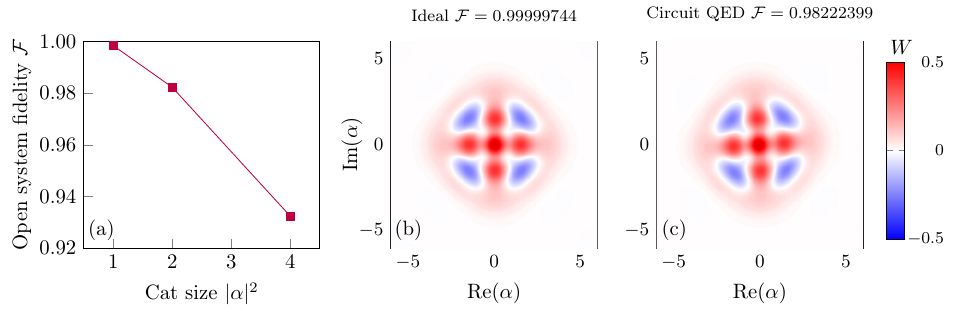}% Here is how to import EPS art
\caption{\label{fig:CatWignerFuncs} {{Preparation of a four-component cat state with a circuit QED system consisting of a tunable transmon coupled to an LC resonator via an asymmetric SQUID in the presence of qubit and resonator energy relaxation and dephasing. (a) The open system fidelity for different cat sizes decreases, as expected, with cat size. Wigner functions of a four-component cat state with $\alpha=\sqrt{2}$ prepared by (b) an ideal system and (c) by the circuit QED open system.}}} 
\end{figure}

We now study the preparation of two- and four-component cat states (see Table~\ref{tab:StatePrep}) using our multiphoton control protocol introduced in Sec.~\ref{sec:Multi_control}. At steps where the 2JC operation is applied, we rely on Eq.~\eqref{eq:cQEDHamTLAFinal}, while we use Eq.~\eqref{eq:QubitDrive} for the qubit drive steps. We work in the interaction picture and follow the steps in Table~\ref{tab:StatePrep}. The final state is found by numerically integrating the master equation in Eq.~\eqref{eq:LindbladME}. We find that the fidelity of a two-component cat, $\ket{\mathcal{C}^{+,2}_{\alpha}},$ with $\alpha=\sqrt{2}$ is $\mathcal{F} = 0.97972653$ contrasted with an ideal closed system fidelity $\mathcal{F}=0.99999884.$ While, the fidelity of a four-component cat, $\ket{\mathcal{C}^{++,4}_{\alpha}},$ with $\alpha=\sqrt{2}$ is $\mathcal{F} = 0.98222399$ compared to its ideal closed system counterpart $\mathcal{F} = 0.99999744.$ 

{Figure ~\ref{fig:CatWignerFuncs} shows that the open system fidelity substantially degrades as a function of cat size, which roughly captures the circuit depth needed to prepare the target state.} Figure~\ref{fig:CatWignerFuncs}(b) shows the Wigner function of $\ket{\mathcal{C}_\alpha^{++,4}}$ with $\alpha=\sqrt{2}$ when the state is prepared using an ideal qubit-oscillator system (as in Sec.~\ref{sec:Multi_control}). Figure~\ref{fig:CatWignerFuncs}(c) shows the Wigner function of the same target state prepared with a circuit QED open system. The state prepared in the presence of spurious terms and decoherence remarkably maintains a high fidelity to the target state, even though it is visually clear that it is off by a slight displacement and rotation. This suggests that with some additional corrective displacement and rotation operations, either in between operations or at the end, the final state fidelity in a realistic open system can be further boosted. 

The circuit QED implementation above can similarly be extended for the multiple oscillator state generation protocol outlined in Sec.~\ref{sec:Multimode}. The system consisting of one qubit and two oscillators would employ a single tunable transmon qubit coupled to two LC resonators each via a distinct asymmetric SQUID (two SQUID couplers in total). To perform a two-photon operation on one of the resonators, the qubit is tuned to that particular resonator's two-photon resonance, resulting in a Hamiltonian similar to the one presented in Eq. \eqref{eq:cQEDHamTLAFinal}. The qubit and resonator will have minimal interactions with the other resonator as they will be far-detuned. The necessary selective qubit rotations for a given state can be implemented by driving the qubit in the dispersive regime at the frequency given in Eq. \eqref{eq:MultiOscillatorDispersiveShift}. {The sequential structure of the multi-oscillator protocol, where a multiphoton operation is applied to one oscillator at a time, and reliance on additional qubit-dependent operations will be more negatively affected by the auxiliary qubit decay channel than the single-oscillator case.} We leave the in-depth analysis of the multi-oscillator implementation using superconducting circuits as a future direction, where a quantitative analysis of the regimes of two-photon operation and the effects of cross-talk will be conducted.

{\subsection{Mitigating oscillator and ancilla decoherence shot-to-shot}}

{ In a given time-evolution period, the system interacts with the environment and all possible decoherence effects can occur with some probability. To better understand the effect of each error on the state preparation, we can isolate each error and analyze its impact depending on where in the state preparation circuit it occurs, e.g. during or after an $n$JC operation vs. during or after a qubit rotation. To better understand the effect of an error on state preparation, we calculate
\begin{align}
    E(\lambda)=\hat{Q}^{(n)}(\lambda\tau)E\hat{Q}^{(n)}((1-\lambda)\tau') \qquad \text{and}\qquad
    \widetilde{E}(\lambda)=\hat{C}(\lambda\tau
')E\hat{C}((1-\lambda)\tau'),
\end{align}
where $E\in \{\hat{\sigma}_x,\hat{\sigma}_z,\aop,\adag\aop\}$ is an error, $\hat{Q}^{(n)}$ is the $n$JC unitary and $\hat{C}$ is the qubit drive. Here, we assume the coupling/drive strength is fixed and simply parameterize the operations with $\lambda\in[0,1]$, which dictates at what point during the gate the error occurs. We note that oscillator errors occurring during qubit rotations do not affect the operation as they only act on the oscillator Hilbert space, whereas qubit errors affect both qubit rotations and $n$JC operations.}

{We first consider the conjugated errors that commute into concrete decipherable operations. A qubit phase flip during an $n$JC gate yields \begin{align}
    \hat{Q}(\lambda \tau)\hat{\sigma}_z\hat{Q}((1-\lambda)\tau)=\hat{Q}(2(\lambda-1)\tau)\hat{\sigma}_z.
\end{align}
This means that the phase flip re-parametrizes the gate time to the effective duration $(2\lambda-1)\tau$ and leaves a residual phase flip. Meanwhile, oscillator dephasing during an $n$JC gate yields \begin{align}
    \hat{Q}(\lambda \tau)\adag\aop\hat{Q}((1-\lambda)\tau)=\adag\aop \hat{Q}(\tau)+\frac{n}{2}\left(\hat{Q}(-\tau)-\hat{Q}(2(\lambda -1)\tau)\right)\hat{\sigma}_z.
\end{align}
Interestingly, oscillator dephasing conjugates to an $n$JC operation followed by a dephasing event superposed with a qubit dephasing followed by a superposition of differently re-parametrized $n$JC operations.
The other errors during an $n$JC or qubit drive operation do not collapse to a concrete decipherable operation. A photon loss or gain event is detectable with a parity mod $n$ measurement since the oscillator occupancy changes from $\mathcal{H}_{k,n}\rightarrow\mathcal{H}_{k\pm1,n}.$ Qubit bit flip errors are more tricky and elusive since the oscillator remains in the same subspace, and depending on the bosonic code used, this could result in a logical, undetectable and uncorrectable error.}

{To mitigate ancilla decay and dephasing shot-to-shot, qubit resets can be added at the end of each pair of $n$JC and qubit rotation operations. One could also consider substituting the transmon ancilla with a noise-biased ancilla in which bit-flips are suppressed such as the Kerr-cat qubit \cite{Ding_QCtrlKerrCat_2025}. Additionally, oscillator modular parity measurements using the ancilla can be implemented at the end of each pair of operations, which are standard in rotationally symmetric bosonic codes \cite{Grimsmo_RotSymm_BosonicQEC_2020}. While these suggestions render the state preparation procedure probabilistic and preparation time longer, they are standard components in most fault-tolerant bosonic architectures \cite{LachanceQuirion_AutonomousGKPQEC_2023,Lemonde_HardwareEfficientBosonicPerspective_2024,Chen_FTCatPrep_2026}. Such a change would require altering and generalizing the $n$LE protocol presented here to work with qubit resets and is an interesting avenue for future research. Another interesting direction is to generalize the $n$LE protocol to a path-independent control protocol \cite{Ma_PathIndepGates_2020,Ma_PathIndepCtrl_2022}. This can be done by utilizing more levels in the ancilla such that when decay errors occur they are flagged and the outcome is either discarded or a mitigation operation is developed for it, e.g. based on the error conjugation for applicable errors such as those described above.}

{\subsection{Limitations on multiphoton qubit-oscillator interactions and selective qubit rotations}}
{The substantial improvements in state preparation times presented earlier in this paper  are hindered by some practical limitations. We discuss three main limitations to the protocols discussed in this paper. The first limitation concerns sideband control (linear and nonlinear) due to the shorter gate time for higher Fock states. The second limitation is strictly related to the multiphoton $n$JC operation and validity of the effective Hamiltonian generating the operation for larger circuit depths. The third limitation concerns the selective qubit rotations used in the FTP and generalized multi-oscillator $n$LE protocols.}

{There are two main device architectures for implementing sideband controls. One relies on tuning the auxiliary qubit on- and off-resonance from the desired sideband interaction; the $n$JC interaction is generated by a nonlinear coupler (possibly with a fixed bias such as in Sec.~\ref{sec:cQED}) \cite{Ayyash_DrivenTwoPhoton_2024}. The other approach relies on parametrically driving a nonlinear coupler, which results in a mediated resonance \cite{Lemonde_HardwareEfficientBosonicPerspective_2024}. In both cases, the "switching" of the interaction to be on and off for the gate duration is usually done in an adiabatic manner such that diabatic transitions are not excited and dephasing errors are minimized \cite{Hofheinz_SynthesisExp_2009}. The shorter the gate duration gets, the more rapid the switching of the interaction has to be (for both approaches). For larger Fock cutoffs, the gate duration becomes short enough that it is no longer possible to adiabatically tune the qubit or drive the coupler, and as a result, adverse effects such as dephasing and coupler heating are amplified. Furthermore, the tuning of the qubit or coupler, in this case, behaves almost like a step function, which has high-frequency components that can excite higher-order multiphoton resonances \cite{Dai_DUSTSpectroscopy_2025}. Following Eq.~\eqref{eq:MultiphotonTimeEstimate}, the upper bound for the $n$JC gate time for the swap $\ket{e}\ket{(j-1)n}\leftrightarrow\ket{g}\ket{jn}$ is $$T_{\text{swap}}=\frac{\pi}{|g_n|\sqrt{\frac{(jn)!}{((j-1)n)!}}}.$$ Then, using $n=2,\,|g_2|=2\pi\times25\text{MHz}$ and $j=10$, we get $T_{\text{swap}}\approx1\text{ ns}$, which is difficult to achieve with current technology and sets a rough cutoff for this parameter set. Similarly, for $n=1,\, |g_1|=2\pi\times 100\text{MHz}$ and $j=21,$ we get $T_{\text{swap}}\approx1.1 \text{ ns}.$ Thus, the Fock cutoff for which sideband controls remain advantageous while dependent on the system parameters, will be limited to below $\sim 20$ photons.}

{Another limitation concerning the multiphoton sideband controls is that of the rotating-wave approximation (RWA). As discussed above in the circuit QED section, realistic device interactions do not naturally occur in the $n$JC form we desire. The RWA validity depends on parameter regimes and timescale considered, as well as energy occupancy of the qubit-oscillator (see for example \cite{Ayyash_MultiphotonDispersive_2025}). Thus, another bound on the tolerable Fock cutoff stems from the validity of the RWA. The longer circuit depths also mean longer evolution timescales that accumulate deviations from the RWA Hamiltonian assumed. As an example, consider the interaction Hamiltonian arising from a realistic coupling in the interaction picture assuming two-photon resonance between the qubit and oscillator$$\hat{H}_{2}=g_2(\hat{\sigma}_+\aop^2 +\sigM\adagT + e^{i(\omega_q+2\omega_o)t}\sigP\adagT+ e^{-i(\omega_q+2\omega_o)t}\sigM\aop^2).
$$Under the RWA, this Hamiltonian simplifies to the 2JC Hamiltonian $$
    \hat{H}_{2} \overset{\text{RWA}}{\simeq} \hat{H}_{2\text{JC}}=  g_2(\sigP \aop^2 + \sigM \adagT).$$ We set the initial state to be $\ket{e}\ket{N},$ and we evaluate the fidelity of the initial state evolved under the non-RWA Hamiltonian compared to evolving under the 2JC Hamiltonian, namely we compute $$\mathcal{F}_{\text{RWA}}(t)=|\bra{e}\bra{N} e^{i \hat{H}_2 t}e^{-i\hat{H}_{2\text{JC}}t}\ket{e}\ket{N}|^2.$$ Assuming the Hamiltonian parameters in Table~\ref{tab:CircuitQEDTable}, for $N=20$ and $t=2\pi/g_2$, we find $\mathcal{F}_{\text{RWA}}=0.989491.$ This shows the degradation in the 2JC operation for larger Fock states, where the counter-rotating terms have a more tangible effect.}

{The selective qubit rotations require operating in the dispersive regime, which sets a bound on the qubit driving strength that could be used. This makes the operation bottlenecked by the allowable driving bounded by the multiphoton dispersive shift. Another problem is that the spectral resolvability of each Fock state gets worse as the Fock number grows, which makes the accurate driving of the correct transition difficult \cite{Wang_PhotonNumCavity_2021}. This is partly mitigated by using the multiphoton dispersive regime since the frequencies are higher-order polynomials in the Fock number. However, the multiphoton dispersive regime itself is also limited in applicability to specific parameters and qubit-oscillator energy occupancy \cite{Ayyash_MultiphotonDispersive_2025}. Thus, the selective qubit rotations operation will also be limited to lower circuit depths due to these issues.}
\\

\section{Conclusion and Outlook}\label{sec:Conc}

We introduced a multiphoton generalization of the LE protocol~\cite{LawEberely_ArbControl_1996}, which we used to generate rotationally symmetric oscillator states. We found that our protocol provides substantial reductions in state preparation times for states of interest to bosonic quantum error correction such as multicomponent cat, binomial and GKP states. We found the $n$LE protocol to provide substantial improvements over conditional displacement state preparation schemes especially for states with low Fock cutoffs. The improvements enabled by our protocol can significantly improve the performance of bosonic codes on planar superconducting hardware, which is currently limited by shorter coherence times than its three-dimensional counterpart. We additionally performed optimal control calculations that provided further speedups to the $n$LE protocol. Furthermore, we introduced a second protocol, FTP, that enables arbitrary state preparation using a combination of different JC interaction orders and selective qubit rotations. We found that FTP performs better than LE when the state is sparse in its Fock support. The protocols we introduced were then generalized to the case of multiple oscillators coupled to the same auxiliary qubit.

We validated the robustness of our protocol by numerically simulating a realistic circuit QED system with spurious terms in the presence of decoherence for the preparation of a four-component cat state.

{We discussed the practical limitations of using sideband linear and multiphoton control in realistic settings. We found that sideband controls are inherently faster than conditional displacements, but their gate times become too short at larger circuit depths, which restricts the number of photons used. An interesting direction worth further investigating is the combined use of (linear and nonlinear) sideband and conditional displacement controls for state preparation.}

We considered the possibility of using higher-order interactions beyond $n=2$ in App.~\ref{app:HigherOrderControl}. Additionally, we provided examples of state preparation for the case of two oscillators in App.~\ref{app:MultiOsc}.

{We believe that the protocols introduced in this work can serve as an important tool in near-term improvements for implementations of bosonic codes. Overall, this work can aid in a near-term break-even bosonic QEC demonstration on planar superconducting circuits, which has not yet been realized.}

\begin{acknowledgements}
    This paper is dedicated to the late J.H. Eberly and R. Laflamme. We thank Xicheng (Christopher) Xu, Baptiste Royer (and his group), and Tristan Lismer for useful discussions, and Andrii Sokolov for a useful email exchange. NG and MM were supported by the Institute for Quantum Computing through funding provided by the Natural Sciences and Engineering Research Council of Canada. SA was supported by Japan’s Ministry of Education, Culture, Sports, Science and Technology’s Quantum Leap Flagship Program Grant No. JPMXS0120319794 and the Center of Innovations for Sustainable Quantum AI (Grant No.
JPMJPF2221) from the Japan Science and Technology Agency.
\end{acknowledgements}

\section*{Data Availability}
The data that support the findings of this article are not
publicly available. The data are available from the authors
upon reasonable request.

\section*{Author Contributions}
MA conceived the main idea. NG performed analytical and numerical calculations with assistance from MA. SA performed OCT calculations and analysis. SA and MM provided critical input on the analysis and
interpretation of the results. MA and NG co-wrote the manuscript with feedback from SA and MM.

\appendix

\section*{Appendices}

\section{Using Higher-Order Interactions }\label{app:HigherOrderControl}

In the main text, we focused on the use of two-photon interactions for state preparation. In this appendix, we discuss the use of higher-order interactions, namely $n=3$ and $n=4$. We consider their use to prepare states in $\mathcal{H}_{0,n}$. We note that such interactions have been experimentally realized with very weak coupling strengths \cite{Saner_GenArbSuperpositionSqz_2024,Vaneslow_FourPhotonDissip_2025}.
\begin{figure}[t]
\centering
\includegraphics[scale=1,trim={.5cm .4cm 0cm 0cm}]{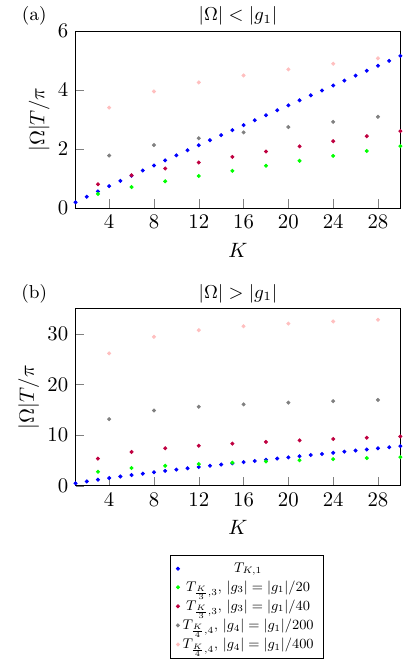}% Here is how to import EPS art
\caption{\label{fig:PrepTimeApp} {Rotationally symmetric state synthesis time for higher-order interactions as a function of number of steps. In $(a)$, $|\Omega|=2\pi\times200\text{ MHz}$, and in $(b)$, $|\Omega|=2\pi\times25\text{ MHz}$. The blue dots represent $n=1$ and are identical to Fig.~\ref{fig:PrepTime}. In the case of $n=3$ ($n=4$), the green (grey) dots have $|g_3|=|g_1|/20$ ($|g_4|=|g_1|/200$), while $|g_3|=|g_1|/40$ ($|g_4|=|g_1|/400$) for the purple (pink) dots. The three-photon (four-photon) time estimates have support exclusively on multiples of three (four) steps, since for each three (four) steps using a linear interaction, we need one step using a three-photon (four-photon) interaction.}} 
\end{figure}
We recall the time estimate to prepare rotationally symmetric states in $\mathcal{H}_{0,n}$ using an $n$-photon interaction in $K$ steps,
\begin{align}\label{eq:MultiphotonTimeEstimateApp}
    T_{K,n}=\frac{K\pi}{|\Omega|}+\sum_{j=1}^K\frac{\pi}{|g_n|\sqrt{\frac{(jn)!}{((j-1)n)!}}}.
\end{align}
Similarly to the case of the two-photon interaction, the number of steps is reduced in the three- and four-photon interaction and the speed up in the Rabi swaps provided by the factor $\sqrt{(jn)!/((j-1)n)!}$ increases. The coupling strengths, though, are even smaller than those of the two-photon interaction. We now compare the time estimates for the three- and four-photon interactions to that of the linear interaction. As in the main text, we consider two regimes, $|\Omega|<|g_1|$ and $|\Omega|>|g_1|.$ We set $|g_1|=2\pi \times 100 \text{ MHz}$ for both cases, and we follow the main text with $|\Omega|=2\pi \times25\text{ MHz}$ for $|\Omega| <|g_1|$ and $|\Omega|=200\text{ MHz}$ for $|\Omega|>|g_1|.$ In Fig.~\ref{fig:PrepTimeApp}(a), we compare the time estimates in the $|\Omega|<|g_1|$ regime for $n=3$ with $|g_3|=|g_1|/20=2\pi\times5\text{ MHz}$ (green dots) and $|g_3|=|g_1|/40=2\pi\times2.5\text{ MHz}$ (purple dots) and $n=4$ with $|g_4|=|g_1|/200=2\pi\times 0.5\text{ MHz}$ (gray dots) and $|g_4|=|g_1|/400=2\pi\times 0.25 \text{ MHz}.$ We observe that the gradient of the time estimates decreases with increasing interaction order. However, with diminishing coupling strengths, the starting point for the time estimate becomes larger and, thus, affects where the multiphoton time estimates cross the linear time estimate, i.e., where they become better. The same reasoning similarly applies to the arbitrary state preparation problem using a combination of interaction orders beyond $n=2.$ In the $|\Omega|>|g_1|$ regime, shown in Fig.~\ref{fig:PrepTimeApp}, we find the higher-order interaction generally perform worse than the linear interaction with fourth-order interactions requiring much more than ten times the preparation time of the linear interaction. Thus, when the qubit driving is allowed to be strong, the linear interaction reigns supreme for most cases. However, as noted in the main text, generally, it is preferable in an experimental setting to minimize qubit driving strengths to avoid spurious qubit excitations.

\section{Multi-Oscillator Examples }\label{app:MultiOsc}

In this appendix, we provide time estimates for rotationally symmetric and arbitrary multi-oscillator states, specifically for two oscillators. The upper bound for the total state generation time in $\mathcal{H}_{0,n_1}\otimes\mathcal{H}_{0,n_2}$ is  
\begin{align}\label{eq:time_estimate_AppB}
    T_{\substack{L_1,n_1;\\L_2,n_2}}=&(L_1 + (L_1 + 1)L_2)\frac{\pi}{|\Omega|} +\sum_{j=1}^{L_1}\frac{\pi}{|g_{n_1}|\sqrt{\frac{(jn_1)!}{((j-1)n_1)!}}} +(L_1 + 1)\sum_{j=1}^{L_2}\frac{\pi}{|g_{n_2}|\sqrt{\frac{(jn_2)!}{((j-1)n_2)!}}}.
\end{align}
Time estimates for states with support over different $\mathcal{H}_{k_1,n_1}\otimes\mathcal{H}_{k_2,n_2}$ subspaces are found by summing the generalized heights over the multidimensional punch card introduced earlier. Thus, the multi-oscillator FTP generalization time estimate reads as
\begin{align}\label{eq:FTPTimeMultiOsc}
    T_{\substack{L_1,n_1;\\L_2,n_2}}^\text{FTP}=T_{\text{b}}+T_{\text{c}_1}^{(n_1)}+T_{\text{c}_2}^{(n_2)},
\end{align}
where
\begin{align}
    T_{\text{c}_1}^{(n_1)}=\sum_{k_2=0}^{n_2-1}\sum_{k_1=0}^{n_1-1} 
    \Biggl( 
    h_{k_1,k_2}^{(n_1,n_2)}\frac{\pi}{|\Omega|}+\sum_{j=1}^{h_{k_1,k_2}^{(n_1,n_2)}}\frac{\pi}{|g_{n_1}|\sqrt{\frac{(jn_1+k_1)!}{((j-1)n_1+k_1)!}}}
    \Biggr)
\end{align}
and
\begin{align}
    T_{\text{c}_2}^{(n_2)}=\sum_{k_1=0}^{L_1}\sum_{k_2=0}^{n_2-1} \Biggl( \mathfrak{h}_{k_1,k_2}^{(n_1,n_2)}\frac{\pi}{|\Omega|}+\sum_{j=1}^{\mathfrak{h}_{k_1,k_2}^{(n_1,n_2)}}\frac{\pi}{|g_{n_2}|\sqrt{\frac{(jn_2+k_2)!}{((j-1)n_2+k_2)!}}}\Biggr) .
\end{align}
The time estimate for generating the base state $T_b$ assumes a linear interaction, which can be simply found using Eq. \eqref{eq:time_estimate_AppB} and setting $n_1=n_2=1$. For the time estimates given below, we assume that the qubit drive and one- and two-photon interaction couplings are the same as the main text ($|\Omega|=2\pi \times 25 \text{ MHz}$, $|g_1|=2\pi \times 100 \text{ MHz}$, and $|g_2|=2\pi \times 25 \text{ MHz}$). 

\subsection{Rotationally symmetric multi-oscillator state preparation in $\mathcal{H}_{0,n_1}\otimes\mathcal{H}_{0,n_2}$}

An example of a rotationally symmetric state in $\mathcal{H}_{0,n_1}\otimes\mathcal{H}_{0,n_2}$ is the NOON state $\ket{\psi_t} = (\ket{N,0}) + \ket{0,N})/\sqrt{2}$, when $n_1 = n_2 = 2$ and $N$ is an even integer. In Table~\ref{tab:NOON}, the coupling parameters used to synthesize a NOON state with $N=2$ at each step are shown. Along with the parameters, the closed system fidelities, obtained through simulating the state synthesis protocols, are given for the linear and multiphoton state generations. To generate the NOON state using linear interactions, 4 steps (8 operations) were needed resulting in a time estimate of $$T_{\substack{L_1=2,n_1=1;\\L_2=2,n_2=1}}^{\ket{\text{NOON}}}\lessapprox97\text{ ns}.$$ In comparison, for the two-photon interaction, the state generation took 2 steps (4 operations), yielding a time estimate of $$T_{\substack{L_1=1,n_1=2;\\L_2=1,n_2=2}}^{\ket{\text{NOON}}}\lessapprox68\text{ ns}.$$ Thus, the time estimates indicate a speedup of $30\%$ for the multiphoton protocol. The non-unital fidelities presented in Table~\ref{tab:NOON} are a result of spurious relative phase accumulations on the qubit attributed to the dispersive shift, and non-resonant qubit excitations in the selective qubit driving that occurred during the state generation simulation. These spurious effects could easily be corrected with additional qubit phase corrections \cite{Sharma_StateSynthsisRev_2016}.

\renewcommand{\arraystretch}{1.2}
\begin{table}[t]

\centering
\resizebox{10cm}{!}{%

\begin{tabular}{ccc}
\textbf{Operation} & \textbf{Parameters} & \textbf{Quantum State} \\ \hline
\multicolumn{3}{c}{$n_1=n_2=1$,  $\mathcal{F} = 0.99333450$} \\
\hline
$\hat{C}^{\ket{0,0}}_1$ & $ \Omega_1 \tau = \pi/4,\ \omega_d = \omega^{(1,1)}_{0,0}$ & $\hspace{-1.2mm} -i|e,0,0\rangle + |g,0,0\rangle$ \\ 
$\hat{Q}^{(1)}_{1}$ & $g_{n_1} \tau = \pi/2$ & $- |g,1,0\rangle + |g,0,0\rangle $ \\ 
$\hat{C}^{\ket{1,0}}_{2}$ & $\Omega_{2} \tau = \pi/2,\ \omega_d = \omega^{(1,1)}_{1,0}$ & $\hspace{1.4mm}i|e,1,0\rangle+|g,0,0\rangle $  \\ 
$\hat{Q}^{(1)}_{2}$ & $g_{n_1} \tau = \pi/2\sqrt{2}$ & $\hspace{2.55mm}|g,2,0\rangle + |g,0,0\rangle$ \\ 

$\hat{C}^{\ket{0,0}}_{3}$ & $\Omega_{3} \tau = \pi/2,\ \omega_d = \omega^{(1,1)}_{0,0}$ & $\hspace{2.55mm}|g,2,0\rangle -i |e,0,0\rangle$  \\ 
$\hat{\mathfrak{Q}}^{(1)}_{1}$ & $g_{n_2} \tau = \pi/2$ & $\hspace{2.55mm}|g,2,0\rangle - |g,0,1\rangle$ \\ 
$\hat{C}^{\ket{0,1}}_{4}$ & $\Omega_{4} \tau = \pi/2,\ \omega_d = \omega^{(1,1)}_{0,1}$ & $\hspace{2.55mm}|g,2,0\rangle + i|e,0,1\rangle$  \\ 
$\hat{\mathfrak{Q}}^{(1)}_{2}$ & $g_{n_2} \tau = \pi/2\sqrt{2}$ & $\hspace{2.55mm}|g,2,0\rangle + |g,0,2\rangle$ \\ 
\hline
\multicolumn{3}{c}{$n_1=n_2=2$, $\mathcal{F} = 0.99690060$} \\
\hline
$\hat{C}^{\ket{0,0}}_{1}$ & $ \Omega_{1} \tau = \pi/4,\ \omega_d = \omega^{(2,2)}_{0,0}$ & $\hspace{-1.15mm}-i|e,0,0\rangle + |g,0,0\rangle$ \\ 
$\hat{Q}^{(2)}_{1}$ & $g_{n_1} \tau = \pi/2\sqrt{2}$ & $- |g,2,0\rangle + |g,0,0\rangle $ \\ 
$\hat{C}^{\ket{0,0}}_{2}$ & $\Omega_{2} \tau = \pi/2,\ \omega_d = \omega^{(2,2)}_{0,0}$ & $-|g,2,0\rangle-i|e,0,0\rangle $  \\ 
$\hat{\mathfrak{Q}}^{(2)}_{1}$ & $g_{n_2} \tau = \pi/2\sqrt{2}$ & $-|g,2,0\rangle - |g,0,2\rangle$ \\ 

\end{tabular}
 }
\caption{Linear and multiphoton interaction parameters for NOON state synthesis. The quantum states shown are the result after applying the corresponding operation, where we originally start in the combined system's ground state.}
\label{tab:NOON}
\end{table}
\subsection{Arbitrary multi-oscillator state preparation}

We now consider arbitrary state preparation for the case of two oscillators. Let the target state be the entangled two-mode Bell-cat state
\begin{align}
    \ket{\mathcal{C}^{+,2}_{\alpha_1, \alpha_2}}=\frac{1}{\mathcal{N}}(\ket{\alpha_1, \alpha_2}+\ket{-\alpha_1, -\alpha_2})= \frac{1}{\mathcal{N}} \sum_{m,n=0}^{\infty} \frac{\alpha_1^m \alpha_2^n}{\sqrt{m!\,n!}} 
\Big[ 1 + (-1)^{m+n} \Big] \ket{ m, n }, 
\end{align}
whose Fock population comprises states with strictly even total photon number across both oscillators. For estimating the number of steps and synthesis time, we assume $\alpha_1 = \alpha_2 = \sqrt{2}$ and a truncation at 10 photons for each oscillator. This truncation would result in a target state with fidelity greater than $0.99999$ under perfect unitary state synthesis. The one-photon interaction protocol requires 115 steps (230 operations) and has a time estimate of $$T_{\substack{L_1=10,n_1=1;\\L_2=105,n_2=1}}^{\ket{\mathcal{C}_{\alpha_1, \alpha_2}^{+,2}}}\lessapprox2.59\mu\text{s}.$$ Conversely, using two-photon interactions necessitates 61 steps (122 operations) that bounds the synthesis time to 
\begin{equation}
    T_{\substack{L_1=10,n_1=2;\\L_2=10,n_2=2}}^\text{FTP}\lessapprox1.54\mu\text{s},\nonumber 
\end{equation}\vspace{0.1cm}

\noindent exemplifying a $41\%$ faster state preparation time. The number of steps and time quoted for the multiphoton protocol includes generating the base state $\ket{\psi_b} = c_1\ket{0,0} + c_2\ket{1,1}$ using a linear interaction. 

Now consider the example where the two-oscillator target state is not sparsely populated, such as $\ket{\psi_t}=\sum_{k_1,k_2=0}^{4}\ket{k_1,k_2}/\sqrt{25}.$ The one-photon interaction synthesis procedure needs 24 steps (48 operations) and has a time estimate of $$T_{\substack{L_1=4,n_1=1;\\L_2=20,n_2=1}}^{\ket{\psi_t}}\lessapprox564\text{ns}.$$ 
In the case of using two-photon interactions, the generation protocol also demands 24 steps (48 operations), but the time estimate is $$T_{\substack{L_1=4,n_1=2;\\L_2=4,n_2=2}}^\text{FTP}\lessapprox692\text{ns},$$
which includes the steps required to generate the base state $\ket{\psi_b} = \sum^1_{k_1,k_2=0} c_{k_1,k_2}\ket{k_1, k_2}$ using a one-photon process. 
As expected, the two-oscillator FTP does not offer a speedup when the target state is not sparsely populated.
\bibliographystyle{quantum}
\bibliography{references}

\end{document}